\documentclass[aps,prx,twocolumn,groupedaddress,superscriptaddress,longbibliography,nofootinbib]{revtex4-1}

\usepackage{amsmath,amssymb}
\usepackage{cancel}
\usepackage{graphicx}
\usepackage{multirow}
\usepackage{bm}
\usepackage{mathtools}
\usepackage{dsfont}
\usepackage{amsfonts}
\usepackage{xcolor}
\usepackage[normalem]{ulem}
\usepackage{url}
\usepackage{wasysym}
\usepackage{cancel}

\usepackage[colorlinks=true,linkcolor=magenta,citecolor=magenta]{hyperref}

\def\ba#1\ea{\begin{align} #1\end{align}}
\def\bg#1\eg{\begin{gather}#1\end{gather}}
\def\bpm{\begin{pmatrix}}
\def\epm{\end{pmatrix}}
\newcommand{\nn}{\nonumber}

\renewcommand{\b}[1]{{\boldsymbol #1}}

\newcommand{\bk}{\b k}

\newcommand{\mf}[1]{\mathfrak{#1}}
\newcommand{\der}{\partial}

\newcommand{\ket}[1]{| #1 \rangle}
\newcommand{\bra}[1]{\langle #1 |}
\newcommand{\brk}[2]{\langle #1 | #2 \rangle}

\newcommand{\eq}[1]{Eq.~\eqref{#1}}
\newcommand{\fig}[1]{Fig.~\ref{#1}}

\newcommand{\ourtitle}{Quantum Geometry and Landau Levels of Quadratic Band Crossings}

\newcommand{\magenta}[1]{\textcolor{magenta}{#1}}

\begin{document}
\title{\ourtitle}

\author{Junseo \surname{Jung}}
\affiliation{Center for Correlated Electron Systems, Institute for Basic Science (IBS), Seoul 08826, Korea}
\affiliation{Department of Physics and Astronomy, Seoul National University, Seoul 08826, Korea}
\affiliation{Center for Theoretical Physics (CTP), Seoul National University, Seoul 08826, Korea}

\author{Hyeongmuk \surname{Lim}}
\affiliation{Center for Correlated Electron Systems, Institute for Basic Science (IBS), Seoul 08826, Korea}
\affiliation{Department of Physics and Astronomy, Seoul National University, Seoul 08826, Korea}
\affiliation{Center for Theoretical Physics (CTP), Seoul National University, Seoul 08826, Korea}

\author{Bohm-Jung \surname{Yang}}
\email{bjyang@snu.ac.kr}
\affiliation{Center for Correlated Electron Systems, Institute for Basic Science (IBS), Seoul 08826, Korea}
\affiliation{Department of Physics and Astronomy, Seoul National University, Seoul 08826, Korea}
\affiliation{Center for Theoretical Physics (CTP), Seoul National University, Seoul 08826, Korea}

\begin{abstract}
We study the relation between the quantum geometry of wave functions and the Landau level (LL) spectrum of two-band Hamiltonians with a quadratic band crossing point (QBCP) in two-dimensions.
By investigating the influence of interband coupling parameters on the wave function geometry of general QBCPs, we demonstrate that the interband coupling parameters can be entirely determined by the projected elliptic image of the wave functions on the Bloch sphere, which can be characterized by three parameters, i.e.,  the major $d_1$ and minor $d_2$ diameters of the ellipse, and one angular parameter $\phi$ describing the orientation of the ellipse. 
These parameters govern the geometric properties of the system such as the Berry phase and modified LL spectra.
Explicitly, by comparing the LL spectra of two quadratic band models with and without interband couplings, we show that 
the product of $d_1$ and $d_2$ determines the constant shift in LL energy while their ratio governs the initial LL energies near a QBCP.
Also, by examining the influence of the rotation and time-reversal symmetries on the wave function geometry, we construct a minimal continuum model which exhibits various wave function geometries. We calculate the LL spectra of this model and discuss how interband couplings give LL structure for dispersive bands as well as nearly flat bands.
\end{abstract}

\maketitle

\section{\label{sec:level1}Introduction}

The quantum geometry of wave functions governs various fundamental physical phenomena ranging from the Aharonov-Bohm effect to the topological phases of matter ~\cite{aharonov1959abeffect,zhang2005top,xiao2010berry,thouless1982top,bernevig2006top,fu2007top,chang2008top}.
For instance, it is well-established that the Berry phase~\cite{berry1984quantal,zak1989berry,xiao2010berry,vanderbilt1993electric} and related higher-order geometric quantities play a critical role in describing the Landau level (LL) spectrum of metals, which is compactly formulated in the form of Onsager's semiclassical quantization condition and its generalization~\cite{onsager1952onsager, roth1966onsager, mikitik1999onsager, gao2017onsager, fuchs2018onsager}.

Recent studies of two-dimensional systems with band crossing points~\cite{rhim2019classification, rhim2020quantum, jung2021flatgeo, hwang2021landau, hwang2021flatgeo} have shown that the quantum geometric tensor or related geometric quantities possess more detailed information about the geometric properties of band crossing points than just the Berry phase itself.
For instance, it was shown that the Berry phase associated with the band crossing point is given by the maximum value of the quantum distance~\cite{jung2021flatgeo}. Moreover, in systems where a flat band is touching with a parabolic band quadratically, the flat band exhibits anomalous LL spreading~\cite{rhim2019classification, rhim2020quantum}, which cannot be explained by a simple extension of the Onsager's semiclassical quantization rule. It was shown that the maximum quantum distance around the band crossing point determines the total spreading of the flat band LLs, which in turn induces logarithmic magnetic field dependence of orbital magnetic susceptibility~\cite{rhim2020quantum}.

In this work, we study the quantum geometry of general two-band Hamiltonians with a quadratic band crossing point (QBCP) ~\cite{uebelacker2011qbcp,fang2012qbcp, sun2009topological, chong2008effective} in two-dimensions and its influence on the LL spectrum, generalizing the previous study of the singular flat band in which the flat band has a quadratic band crossing with another dispersive band~\cite{rhim2019classification, rhim2020quantum}. In particular, we aim at revealing the explicit mapping between the interband coupling terms in generic two-band Hamiltonian with QBCPs and the related wave function geometry.
Contrary to the singular flat band systems whose interband coupling is completely described by the maximal quantum distance~\cite{rhim2020quantum,jung2021flatgeo}, the wave function geometry of generic quadratic band crossing Hamiltonian is more complicated, and their geometric structure has yet to be studied. 

We show that among the nine parameters in a general two-band continuum Hamiltonian describing a QBCP~\cite{rhim2020singular,oshikawa1994qbcp}, six correspond to the mass tensors of the two dispersive quadratic bands, while the other three describe the interband coupling. In particular, we find that the wave function geometry of the system in momentum space appears in the form of an elliptic shape on the Bloch sphere, and the three interband coupling parameters determine the major $d_1$ and minor $d_2$ diameters of the elliptic curve and the orientation of the ellipse represented by an angular variable $\phi$. Moreover, we show that the flat band condition is nothing but the condition that the energy eigenvalues of the two-band Hamiltonian have a quadratic analytic form. Under such quadratic form condition, the generic two-band Hamiltonian has only one interband coupling parameter, which is equivalent to the maximum quantum distance, and the corresponding wave function trajectory on the Bloch sphere is a circle whose diameter is equal to the maximum quantum distance, consistent with Refs.~\cite{rhim2020quantum,jung2021flatgeo}.

\begin{figure}[t]
    \centering
    \includegraphics[width=0.5\textwidth]{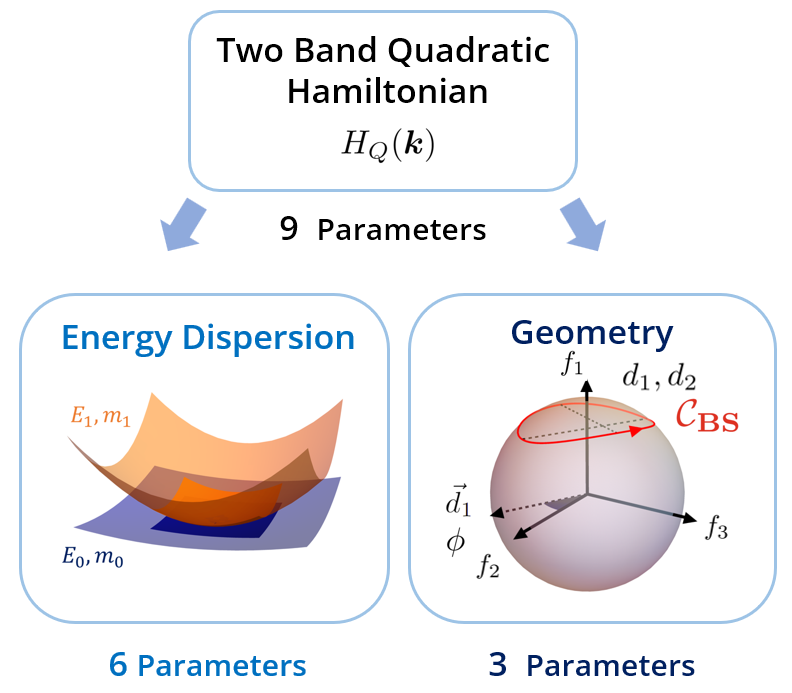}
    \caption{Schematic figures explaning the physical meaning of the nine parameters of a general two-band Hamiltonian with a QBCP. Among these, six are related to the mass tensors of the energy dispersion $E_1, E_2$. The rest describe the shape of the wave functions projected to the Bloch sphere, which generally takes the form of an elliptic curve on the Bloch sphere. $d_1, d_2$ represent the diameter of the two axes of the elliptic curve while $\phi$ indicates how much the axis of the ellipse deviates from the $f_2$-axis.   
}
    \label{Fig1}
\end{figure}

Next, we analyze the LL problem of two-band quadratic Hamiltonians with QBCPs and reveal the role of the interband coupling in the LL spectrum. 
In particular, we compare two quadratic bands with identical mass tensors but different interband couplings to demonstrate how the geometric parameters affect the LL structure. 
Explicitly, we show that the product of two distance parameters, $d_1, d_2$, affects the constant energy shift of LLs, while their ratio controls the energy levels of the initial LLs near the QBCPs. 
Furthermore, we analyze the quantum geometric tensor of the relevant LL wave functions and show how the interband coupling affects the quantum metric distribution of the LL wave functions. 

Finally, by examining the influence of rotation and time-reversal symmetries on the wave function geometry of QBCP Hamiltonians, we propose a minimal continuum model which exhibits various quantum geometry of wave functions. We calculate the LL spectra for this model and observe how varying parameters of this model change the mass tensors and anisotropy ratio $d_1/d_2$, resulting in different LL energy spacings and the relative positions of the LLs.

The rest of the paper is organized as follows. 
In Sec. \ref{sec:QBCP}, we describe the general form of two-band continuum Hamiltonians with QBCPs and explain their geometric properties. 
In Sec.~\ref{sec:Flat}, we show how the flat band condition reduces the geometric parameters using the property of quadratic forms. 
In Sec. \ref{sec:LL}, we discuss how the geometric parameters affect the LL energies and their wavefunctions. 
In Sec. \ref{sec:Sym}, we review how lattice symmetries affect the wave function's geometry. We analyze the results by focusing on n-fold rotation $C_n$ symmetry and time-reversal symmetry, and construct a minimal model demonstrating various wave function geometry. Detailed calculations and derivations are presented in the supplementary materials (SM).

\section{Continuum Model of QBCPs and its geometric properties}\label{sec:QBCP}
In this section, we discuss the set of independent parameters characterizing a generic QBCP. 
After deleting redundant variables by appropriate unitary transformations, 
we will see that nine independent parameters determine the continuum Hamiltonian describing a QBCP. We aim to introduce all of the defining parameters of the QBCP Hamiltonians and organize them into (i) the mass tensors, which describe the local energy dispersion near the BCP, and (ii) the quantum geometric distance and angle parameters, which characterize the interband coupling of the two bands.
A schematic figure summarizing the physical meaning of the parameters in the Hamiltonian is shown in \fig{Fig1}.

Before explaining the details of QBCPs, 
we review a single band model in two dimensions. A general single band model with quadratic dispersion can be written as $H(\bk)=t_1k_x^2+2t_2k_xk_y+t_3k_y^2,$ where $(k_x,k_y)$ are the momenta and $t_1, t_2, t_3$ are constant parameters. For a single band model, $H(\bk)$ corresponds to the energy dispersion, $E(\bk)$, and $t_i$'s naturally become the inverse of the mass tensors, $(2m^{xx})^{-1},( 2m^{xy})^{-1}, (2m^{yy})^{-1}$, respectively. Therefore, the three mass tensors are needed to characterize a general one-band model fully. 

Now, we move to define the parameters of two-band models. A generic two-band Hamiltonian with a QBCP at $\bk=(0,0)$ in 2D is written in the following form,
\ba
    H^{(0)}_{Q}(\bk)=\sum_{a=0}^2\sum_{i=0}^3v_{a,m}k_x^mk_y^{2-m}\sigma_a,
    \label{eq:generic_quad_H}
\ea
where the Pauli matrices $\sigma_{x,y,z}$ describe the two bands and $v_{a,m}$ are constant parameters. Compared to the three mass tensors of the single-band model, twelve parameters, $v_{a,m}$, define the two-band model. However, it is possible to reduce the number of parameters using unitary transformations. Two Hamiltonians, connected by unitary transformation, are considered equivalent Hamiltonians since they have the same energy dispersion and unitarily identical wave functions. Therefore, we apply consecutive unitary transformation to \eq{eq:generic_quad_H} to make the Hamiltonian more concise as in Ref.~\cite{jung2021flatgeo}. After the unitary transformation, the Hamiltonian becomes
\ba
    H_{Q}(\bk)=&[q_1(k_x^2+k_y^2)+q_2(k_x^2-k_y^2)+q_3(2k_xk_y)]\sigma_0 \nn \\
    +&[b_2(k_x^2-k_y^2)+b_3(2k_xk_y)]\sigma_1+[c_3(2k_xk_y)]\sigma_2 \nn \\
    +&[a_1(k_x^2+k_y^2)+a_2(k_x^2-k_y^2)+a_3(2k_xk_y)]\sigma_3. 
    \label{eq:quad_H}
\ea
We note that using three rotation axes of unitary transformations, we reduce three parameters out of the original twelve parameters. As a result, a general two-band continuum model with a QBCP has nine parameters, $q_{1,2,3}, a_{1,2,3}, b_{2,3}$, and $c_3$, which fully define the system. 

Relating it to the one-band model, we assign three mass tensors to each band. Since the coefficients of the Pauli matrices of the Hamiltonian is a quadratic function of momenta, the energy dispersion also scales quadratically. Therefore, we define mass tensors for each band $(i=1,2)$, $m_i^{xx},m_i^{xy},m_i^{yy}$ which describe the energy dispersion. The difference between a one-band model and a two-band model is that there are additional three parameters which do not describe the energy dispersion of the bands. These three parameters will be named interband coupling parameters because they arise due to the coupling of two bands. These coupling parameters depend on the the wave function's singularity around the band crossing point which can be captured by using the notion of the quantum geometry.

\subsection{Mass tensors of the QBCP Hamiltonian}

We first introduce a method of extracting mass tensors of the two-band Hamiltonian. To see this, let us rewrite \eq{eq:quad_H} in a compact form
\ba
    H_{Q}(\bk) = & Q_{0}(k_x,k_y)\sigma_0 + Q_{1}(k_x,k_y)\sigma_1 \nn \\
    + & Q_{2}(k_x,k_y)\sigma_2 + Q_{3}(k_x,k_y)\sigma_3,
    \label{eq:quad_H_compact}
\ea
where $Q_{i}(k_x,k_y)\;(i=0,1,2,3)$ are quadratic functions of $(k_x,k_y)$. 
The energy dispersion of the Hamiltonian in \eq{eq:quad_H_compact} is given by
\ba
    E_{1,2} = Q_0 \pm \sqrt{Q_{1}^{2}+Q_{2}^{2}+Q_{3}^{2}}.
    \label{eq:quad_E}
\ea
Rather than acquiring mass tensors from the energy dispersion of each band $E_{1,2}$, we define $E_{+,-}=\frac{1}{2}(E_{1}\pm E_{2})$, where index $1 (2)$ indicates the energy of the upper (lower) band. This divides $E_{1,2}$ into a smooth function, where elementary definition immediately gives mass tensor, and a non-analytic function, for which we are now going to introduce mass tensor-like quantities. The final mass tensors of $E_{1,2}$ will be a linear combination of the tensors assigned to $E_{+,-}$.

For $E_+$, it is straightforward to calculate the mass tensors. Since $E_+$ is a quadratic expression, the coefficients of each quadratic term become the mass tensor as follows
\ba
    E_{+} &= Q_{0}(k_x,k_y) \nn\\
    &= (q_1+q_2)k_x^2 + 2q_3k_xk_y + (q_1-q_2)k_y^2 \nn\\ 
    &= \frac{1}{2m_{+}^{xx}}k_x^2 + \frac{1}{m_{+}^{xy}}k_xk_y + \frac{1}{2m_{+}^{yy}}k_y^2.
\ea

In contrast, $E_-$ is not a quadratic form in general. $E_-$ has the form of
\ba
E_{-}&=\sqrt{Q_1^2+Q_2^2+Q_3^2} \nn \\
&=\sqrt{\alpha_1k_x^4+\alpha_2k_x^3k_y+\alpha_3k_x^2k_y^2+\alpha_4k_xk_y^3+\alpha_5k_y^4},
\label{eq:Equad}
\ea
where $\alpha_i\;(i=1,\ldots,5)$ are intricate polynomials of the nine parameters in \eq{eq:quad_H}. $E_-$ may reduce to a quadratic expression only when the parameters are fine-tuned. This fine-tuning condition, which we will call the quadratic form condition, reduces the number of parameters of the Hamiltonian and affects the geometry of the band crossing points. However, this fine tuning condition is generally not satisfied and $E_1$ is not expressed as a quadratic form. Therefore, we need to approximate $E_-$ with a quadratic expression $\tilde{E}_-$ to calculate the mass tensors of $E_-$. 

\begin{figure}[t]
    \centering
    \includegraphics[width=0.5\textwidth]{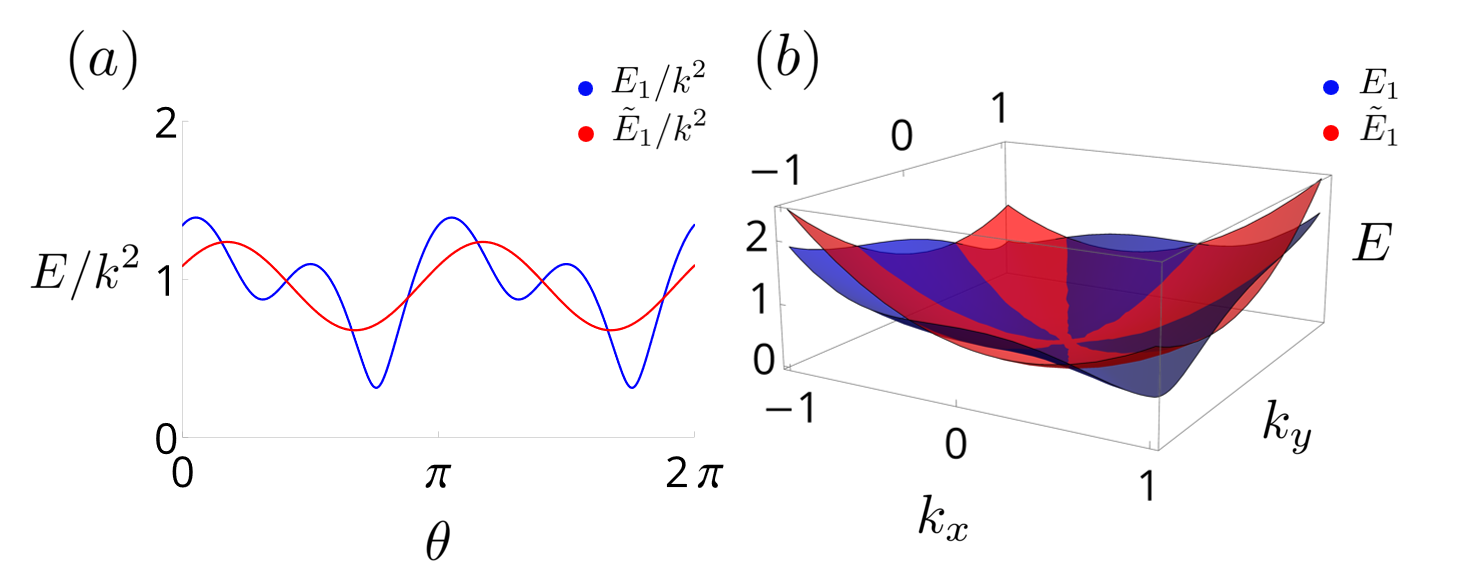}
    \caption{Figures illustrating how $\tilde{E}_-$ approximates the initial energy $E_-$. (a) A graph of the energy divided by $k^2$ as a function of $\theta$ for $E_-$ and $\tilde{E}_-$. (b) The momentum dependence of the energy dispersion. $E_-$ is given by \eq{eq:Equad} with coefficients $\alpha_1=1, \alpha_2=0.3, \alpha_3=0.5, \alpha_4=0.4, \alpha_5=0.2$. $\tilde{E}_-$ is given as \eq{eq:Equadap} with coefficients $\alpha=0.956, \beta=0.129, \gamma=0.248.$}
    \label{Fig:EnergyDisp}
\end{figure}

Since $E_-$ is not differentiable at the origin, naively differentiating twice does not define the mass tensor. Thus, we introduce an approximation via Fourier transform. A quadratic equation $\tilde{E}_-$ can be expressed in the following way, 
\begin{gather}
    \tilde{E}_1=\alpha(k_x^2+k_y^2)+\beta(k_x^2-k_y^2)+\gamma(2k_xk_y), \nn \\
    =k^2(\alpha+\beta\cos{2\theta}+\gamma\sin{2\theta}), \nn \\
    k=\sqrt{k_x^2+k_y^2},
    \label{eq:Equadap}
\end{gather}
where $\theta=\arctan{(k_y/k_x)}$. $E_-$ in \eq{eq:Equad} can also be expressed in the form of $k^2\sqrt{f(\theta)}$, where $\sqrt{f(\theta)}$ is a periodic function of $\theta$ with period $\pi$. Therefore, the best way to approximate $E_-$ using $\tilde{E}_-$ is to define $\alpha, \beta, \gamma$ as the Fourier coefficients of $1, \cos{2\theta}, \sin{2\theta}$ for $E_-$. 
\fig{Fig:EnergyDisp} illustrates an example of extracting mass coefficients of the general quadratic dispersion given by \eq{eq:Equad}. After acquiring $\alpha, \beta, \gamma$ from the Fourier transform, we define $m_{-,xx}=(2\alpha+2\beta)^{-1}, m_{-,xy}=(2\gamma)^{-1}, m_{-,yy}=(2\alpha-2\beta)^{-1}$, as the mass tensors of $E_-$. Combining them with the mass tensors of $E_+$, we obtain the six mass tensors corresponding to the energy dispersion in \eq{eq:quad_E} as follows,
\ba
    m_{1,2}^{xx} = ((m_{+}^{xx})^{-1}\pm (m_{-}^{xx})^{-1})^{-1}, \nn\\ 
    m_{1,2}^{xy} = ((m_{+}^{xy})^{-1}\pm (m_{-}^{xy})^{-1})^{-1}, \nn\\ 
    m_{1,2}^{yy} = ((m_{+}^{yy})^{-1}\pm (m_{-}^{yy})^{-1})^{-1}.
\ea
Thus, we obtained all six parameters of the Hamiltonian associated with the energy dispersion of the two bands.

The mass tensors as defined above do not allow simple analytic expressions in general, since the recipe includes the Fourier transformation of $E_-$. This keeps one from making a clear connection between the quantum geometry of QBCP and its physical properties. However, when the Hamiltonian is nearly isotropic, i.e. $a_1$ is significantly larger than other coefficients $a_{i>1}, b_i, c_i$ in \eq{eq:quad_H}, the calculations are simplified. Since the Landau level problem is exactly solved for an isotropic Hamiltonian, we can apply perturbation method to the Landau level problem for the nearly isotropic regime. Furthermore, in this limit, the Fourier coefficients $\alpha, \beta, \gamma$ of \eq{eq:Equadap} take a simple form. To see that large $a_1$ limit is equivalent to the isotropic limit, note that the term $a_1(k_x^2+k_y^2)\sigma_{3}$ of \eq{eq:quad_H} is the only isotropic contribution to the energy level difference. 

In this large $a_1$ limit, $E_{-}$ is approximated as a quadratic expression, $\tilde{E}_-$, namely
\ba
\tilde{E}_-=a_1k^2(1+\frac{\Delta}{4a_1^2}+\frac{a_2}{a_1}\cos{2\theta}+\frac{a_3}{a_1}\sin{2\theta}),
\label{eq:tildeE}
\ea
where $\Delta=b_2^2+b_3^2+c_3^2$ and the Fourier coefficients of \eq{eq:Equadap} are $\alpha=a_1+\frac{\Delta}{4a_1}, \beta=a_2, \gamma=a_3$. This expression is expanded up to $\mathcal{O}(\delta^2)$, where $\delta$ expresses a ratio between $a_1$ and other coefficients, such as $\delta=a_2/a_1$. We observe that the coefficients $a_2$ and $a_3$ which contribute linearly are diagonal terms of the \eq{eq:quad_H}, while the terms $b_2, b_3$, and $c_3$ which contribute quadratically to the energy are the off-diagonal terms of the Hamiltonian. Thus, we show that the diagonal components contribute mainly to the energy related parameters than the off-diagonal terms. This relation will be reversed for the geometry related parameters, which we will show in the next section. The detailed derivation is provided in the supplementary material \ref{app:Bloch_Sphere}.

\subsection{Interband Coupling Parameters of QBCP Hamiltonians}

In the previous section, we defined the six mass tensors out of nine parameters in \eq{eq:quad_H}, leaving us with three additional parameters. Since the mass tensors describe the band dispersion, it is reasonable to expect that the remaining degrees of freedom describe the behavior of wave functions near the BCP. We organize these free parameters into three geometric measures, namely, the major and minor axis diameters and a rotation angle of an elliptical shape representing the interband coupling of the wave function. These quantities are purely two-band properties that are absent in the single-band Hamiltonian because they are defined based on the nontrivial interband nature of the wave functions. The conventional interband parameters are the off-diagonal terms of the Hamiltonian in \eq{eq:quad_H} involving $b_2, b_3$, and $c_3$. However, our paper introduces coupling parameters based on analyzing the geometry of the wave functions near the BCP. We quantify the interband coupling by using the wave functions and provide a relationship between our interband parameters and the conventional ones.

When defining the interband coupling, we focus on the discontinuity of wave functions around the BCP~\cite{rhim2019classification, rhim2020quantum}. This discontinuity is a crucial feature specific to two-band models. For an isolated band, it is possible to define a gauge such that the wave functions are continuous near each momentum point locally. However, when a BCP exists between two bands, defining a gauge that guarantees continuous wave functions for both bands near the BCP is generally impossible. 

The discontinuity of wave functions can be effectively captured by the Berry phase~\cite{jung2021flatgeo}. Since the Berry phase is gauge-invariant, if there exists a continuous gauge of wave functions in a local region containing the BCP, the Berry phase assigned to a loop around the BCP converges to zero as the loop  can be adiabatically deformed to a point at the BCP. Conversely, if a non-zero Berry phase persists around the BCP, it indicates that wave functions must exhibit discontinuity around the BCP. While the Berry phase serves as a useful metric for detecting singularity in wave functions, further descriptors are required to characterize the entire geometry of the wave functions completely. We aim to define a set of geometric parameters with two distance parameters and one angle parameter, capturing the full features of the wave functions around the BCP.

\begin{figure}[t]
	\centering
	\includegraphics[width=0.5\textwidth]{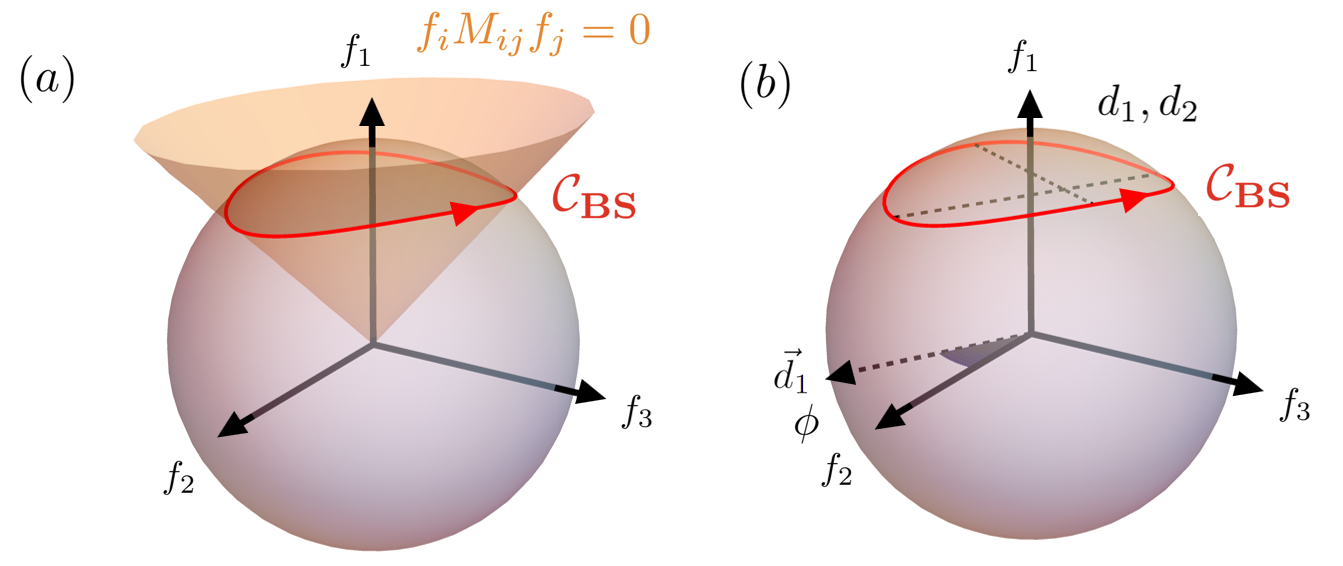}
	\caption{(a) A schematic figure showing how $\mathcal{C}_{BS}$ is given by the intersection of equation $f_iM_{ij}f_j=0$ and the Bloch sphere, where $M_{ij}$ is provided in \eq{eq:Mmatrixo}. The curve formed by the intersection becomes an elliptic shape which can be described by three parameters: $d_1, d_2, \phi$. (b) Definition of the three geometric parameters; $d_1, d_2$ are the major and minor diameters of the elliptic curve and $\phi$ is the angle representing how much major axis, represented as $\vec{d}_1$, deviates from the $f_2$ axis.
	}
	\label{Fig:BlochSphere}
\end{figure}

To characterize the singularity of the wave functions more faithfully, we project the wave functions onto the Bloch sphere. The Bloch sphere is a well-known tool which maps a two-band wave functions to a sphere~\cite{bleu2017bloch,tan2019bloch,graf2021bloch}. We briefly review the concept of Bloch sphere and quantum distance, which is also presented in Ref.~\cite{jung2021flatgeo}. When a two-band Hamiltonian is given as,
\ba
H(\boldsymbol{k})=f_0(\boldsymbol{k})\sigma_0-\boldsymbol{f}(\boldsymbol{k})\cdot\boldsymbol{\sigma},
\label{eq:Ham_fvec}
\ea
the ground state of $H(\boldsymbol{k})$, maps into a point, $\check{\boldsymbol{n}}(\bk)$, on a 2-sphere defined by
\ba
\check{\b n}(\bk)=\frac{1}{2}\frac{\b f(\bk)}{|\b f(\bk)|} \in S^2_{\rm BS},
\label{eq:nvec_def}
\ea
where $S^2_{\rm BS}$ is a sphere with radius, $r=1/2$. 
We note that the ground state of the Hamiltonian does not change when $\bk$ is multiplied by a constant $c$. Since all terms in the Hamiltonian have quadratic order of momentum, $H(c\bk)$ becomes $c^2H(\bk)$, resulting in $\ket{\psi(c\bk)}=\ket{\psi(\bk)}$. 
Therefore, the map, $\check{\b n}(\bk)$, also is invariant under $\bk \rightarrow c\bk$ and $\check{\b n}(\bk)$ is only parametrized by $\theta$, which is the direction of the momentum, $\theta=\arctan{k_y/k_x}$. Thus, the collection of wave functions defined over a finite region in momentum space projected onto the Bloch sphere becomes a loop parameterized as $\check{\b n}(\theta)$, which we call $\mathcal{C}_{BS}$.

The advantage of using the Bloch sphere is that the quantum distance between the wave functions is equal to the geometric distance of the points on the Bloch sphere representing the wave functions. This enables one to relate the geometric objects on Bloch sphere to the properties of the ground state wave function. To be precise, the Hilbert-Schmidt quantum distance measures the distance between two states ~\cite{provost1980riemannian, anandan1990geometry, berry1989quantum,kolodrubetz2017geometry,resta2011insulating,cheng2010quantum} $\ket{\psi(\bk)}, \ket{\psi(\bk')}$ and is given by
\begin{align}
    d^2(\bk,\bk^\prime) = 1-|\brk{\psi(\bk)}{\psi(\bk^\prime)}|^2,
\label{eq:dis_def}
\end{align}
giving a value between zero and one depending on their overlap. This distance is equal to the Euclidean distance $|\check{\b n}(\bk)-\check{\b n}(\bk')|$ of the two points on the Bloch sphere, each representing the two quantum states $\ket{\psi(\bk)}, \ket{\psi(\bk')}$. Therefore, the geometric property of the projected image $\mathcal{C}_{BS}$ of the wave functions on Bloch sphere, such as the shape and size of $\mathcal{C}_{BS}$, is crucial for understanding the ground state properties relevant to the Hilbert-Schmidt quantum distance.

The presence of wave function discontinuity at the BCP can be identified by using the corresponding $\mathcal{C}_{BS}$ on the Bloch sphere, which represents the wave functions around the BCP along a loop encircling the BCP in momentum space. If the wave functions are continuous near the BCP, then as we shrink the loop in momentum space, the wave functions become identical, making $\mathcal{C}_{BS}$ as a point. Conversely, when a singularity exists at the BCP, the wave functions along a momentum space loop do not converge to a point, causing $\mathcal{C}_{BS}$ to be a curve on the Bloch sphere. 

For QBCPs, $\mathcal{C}_{BS}$ is a curve formed by the intersection of an elliptic cone and the Bloch sphere, which is illustrated in \fig{Fig:BlochSphere}. 
The elliptic cone can be obtained from the quadratic equation $\sum_{i,j}f_iM_{ij}f_j=0$ between the three coeffients $f_1, f_2, f_3$ of the Pauli matrices in \eq{eq:Ham_fvec}.
The exact form of $M_{ij}$ and its derivation are in \ref{app:GeoCalc}. Since $\mathcal{C}_{BS}$ is an elliptical curve, it is entirely determined by three parameters $d_1, d_2$, and $\phi$, where $d_1$ and $d_2$ are the major and minor diameters and $\phi$ is the rotation angle of the major diameter $d_1$ with respect to $f_2$ axis when projected to $f_2f_3$-plane. The geometric parameters are illustrated in \fig{Fig:BlochSphere}. We define these geometric parameters to be interband coupling parameters of two bands which captures the essential information governing the quantum geometric properties of the QBCPs.

We calculate the geometric parameters $d_1, d_2$, and $\phi$ of the Hamiltonian in Eq.~(\ref{eq:quad_H}) in the large $a_1$ limit, which corresponds to small $d_1, d_2$ limit. The results are given by
\ba
d_1^2&=\frac{\Delta+\sqrt{\Delta^2-4b_2^2c_3^2}}{2a_1^2}, \nn \\
d_2^2&=\frac{\Delta-\sqrt{\Delta^2-4b_2^2c_3^2}}{2a_1^2}, \nn \\
\phi&=\arctan{\frac{c_3^2-\frac{\Delta}{2}+\frac{\sqrt{\Delta^2-4b_2^2c_3^2}}{2}}{b_3c_3}} \mod \pi,
\label{eq:geopara}
\ea
where $\Delta=b_2^2+b_3^2+c_3^2$ and $b_2>c_3$. When $a_1$ and $b_2c_3$ are fixed parameters, the ratio $d_1/d_2$ is monotonically increasing function of $\Delta$. Thus, the anisotropy of $d_1, d_2$ has one-to-one correspondance with $\Delta$. The detailed calculation is provided in the \ref{app:GeoCalc}.

We note that the coupling parameters, $d_1, d_2$, and $\phi$, are primarily determined by the coefficients $b_2$, $b_3$, and $c_3$, as opposed to the mass tensors which exhibit a greater dependence on $a_2$ and $a_3$, as evident from \eq{eq:tildeE}. This distinction arises from the Hamiltonian $H(\bk)$, where the mass tensors are dominantly influenced by the diagonal terms up to the first order. In comparison, the interband coupling parameters are governed by the off-diagonal terms which correspond to the coefficients of $\sigma_1$ and $\sigma_2$. Therefore, the geometrical parameters are mostly expressed in the conventional interband coupling parameters, $b_2$, $b_3$, and $c_3$, and are independent of the mass tensors.

Using the quantities $d_1$ and $d_2$, we detect the singularity of the BCP as follows. If $\mathcal{C}_{BS}$ turns out to be a point on the Bloch sphere, it implies that the wave functions are identical for every direction of $\theta(=\arctan{k_y/k_x})$. Therefore, the wave functions must be continuous around the BCP, and we define these BCPs to be nonsingular BCP precisely when $d_1=d_2=0$. In contrast, the wave functions are discontinuous at the BCP if and only if $\mathcal{C}_{BS}$ fails to converge to a point and becomes an elliptic loop with $d_1>0$. These BCPs are called singular~\cite{rhim2019classification, rhim2020quantum}. A BCP with $d_1>0, d_2=0$ cannot be detected by the Berry phase since the Berry phase is zero yet the wave function remain discontinuous around the BCP. Therefore, our set of parameters provides a more complete picture about the quantum geometry of
the BCP, than the Berry phase itself.

$d_1, d_2, \phi$, as a whole, fully characterize the geometry of the QBCP so that all other quantities of geometric nature are functions of these parameters. For instance, it was shown in Ref.~\cite{jung2021flatgeo} that the Berry phase can be determined by interband coupling parameters when $d_1=d_2$. This can be further generalized as follows. 
Since the Berry phase equals to, up to sign, one half of the solid angle subtended by $\mathcal{C_{BS}}$, it suffices to give a formula for the solid angle in terms of $d_1, d_2, \phi$, which is given by
\ba
\Phi_{B}(\mathcal{C_{BS}})=\int_{0}^{2\pi}s\sqrt{1-\frac{d_1^2d_2^2}{d_1^2\sin^2{\theta}+d_2^2\cos^2{\theta}}} d\theta \mod 2\pi.
\label{eq:Berryphase}
\ea

Two comments on the derivation of \eq{eq:Berryphase} are in order. First, as $\bk$ runs over a loop enclosing the BCP in the momentum space, its image on the Bloch sphere traces the curve $\mathcal{C}_{BS}$ twice because of the quadratic dispersion, cancelling the aforementioned factor of $\frac{1}{2}$. We check this by observing the invariance of $f_1, f_2, f_3$ under the inversion $\bk\rightarrow-\bk$. Second, depending on the orientation of the curve, the equation's sign changes. If the curve rotates clockwise with respect to the normal vector pointing outward to the surface of the sphere, then the solid angle becomes negative, changing the sign of the Berry phase. Therefore, an additional factor $s$ is added, where $s=\pm1$ depending on the curve's orientation.

\section{Flat band Condition} \label{sec:Flat}
In Ref. \cite{jung2021flatgeo}, it was shown that the flat band condition reduces the number of parameters in the Hamiltonian and restricts $\mathcal{C}_{BS}$ to be a circle. In this section, we introduce a `quadratic form condition', which provide a more general view toward the geometric meaning of the flat band condition in a general setting. Also, we provide an alternate  argument on why the flat band condition implies a circular $\mathcal{C}_{BS}$.

First, we discuss the properties of quadratic forms in two-dimensions, and relate them to QBCP Hamiltonians. Recall that a quadratic form in two variables $(k_x,k_y)$ is a function of the form
\ba
    Q(k_x, k_y)= a k_{x}^{2} + b k_x k_y + c k_{y}^{2} \quad (a,b,c \in \mathbb{R}),
    \label{def.quad}
\ea
where $a,b,c$ are real numbers. Quadratic forms can be added together or multiplied by a scalar, forming a vector space which we denote by $\mathcal{V}$. As can be seen in \eq{def.quad}, we need three scalars to specify a quadratic form, which implies $\dim\mathcal{V}=3$. Therefore, a generic collection of three quadratic forms, $\{Q_1, Q_2, Q_3\}$, constitutes a basis of $\mathcal{V}$, which means that an arbitrary quadratic form $Q$ can be expressed uniquely as $Q = p_1Q_1+p_2Q_2+p_3Q_3$ for some numbers $p_1,p_2,p_3$.

Let us turn to the QBCP Hamiltonian, which has the general form
\begin{gather}
    H_{Q}(\bk) = Q_0(k_x,k_y)\sigma_0 + \sum_{i=1}^{3} Q_i(k_x,k_y)\sigma_{i},
    \label{def.QBCP}
\end{gather}
where the coefficients $Q_i\;(i=0,1,2,3)$ of the Pauli matrices are quadratic forms in $k_x$ and $k_y$. The two energy levels are given by
\ba
    E_{\pm} = Q_0 \pm \sqrt{Q_1^2+Q_2^2+Q_3^2}.
    \label{eq.energy}
\ea
Note that $E_{\pm}$ is not a quadratic form in general, but contains the square root of a quartic function. However, if $E_{-}$ is flat, i.e.,
\begin{gather}
    E_{-} = Q_0 - \sqrt{Q_1^2+Q_2^2+Q_3^2} = 0,
    \label{def.flat}
\end{gather}
the square root of the quartic $Q_1^2+Q_2^2+Q_3^2$ is identical to a quadratic form, $Q_0$. We isolate this observation as a separate condition on the QBCP Hamiltonian: we say that $H(\bk)$ satisfies the quadratic form condition if, in \eq{eq.energy},
\begin{gather}
    Q \equiv \sqrt{Q_1^2+Q_2^2+Q_3^2}\; \text{ is a quadratic form. }
    \label{condition_quad}
\end{gather}
Note that the quadratic form condition is slightly weaker than flatness condition: A Hamiltonian satisfying \eq{condition_quad} may not have a flat band. In that case, however, one can tune one of the bands to be exactly flat by adding an appropriate quadratic form to $Q_0$, without altering the quantum geometry of $H(\bk)$.

Now, we prove that quadratic form condition in \eq{condition_quad} is equivalent to the condition that $\mathcal{C}_{BS}$ is a circle (or a single point). First, we show that quadratic form condition gives circular $\mathcal{C}_{BS}.$ Suppose $Q$ defined in \eq{condition_quad} is quadratic. Then, since $\{Q_1,Q_2,Q_3\}$ span the vector space $\mathcal{V}$, there exist three real numbers $p_1,p_2,p_3$ such that
\begin{gather}
    Q = p_1Q_1 + p_2Q_2 + p_3Q_3.
    \label{eq.basis_expansion}
\end{gather}
Dividing both sides of \eq{eq.basis_expansion} by $Q$, we have
\begin{gather}
    1 = p_1\frac{Q_1}{Q} + p_2\frac{Q_2}{Q} + p_3\frac{Q_3}{Q} = 2\mathbf{p} \cdot \mathbf{\check{n}}(\bk),
    \label{eq.plane}\\ 
    \mathbf{\check{n}} = \frac{1}{2Q}(Q_1,Q_2,Q_3),
    \label{eq.n_vec}
\end{gather}
where $\mathbf{\check{n}}(\bk)$ is the point on the Bloch sphere with radius $1/2$ representing the occupied state at $\bk=(k_x,k_y)$.

Since \eq{eq.plane} is an equation of the plane with normal vector $\mathbf{p}$ and the distance from the origin $1/(2|\mathbf{p}|)$, the image $\mathcal{C}_{BS}$ of $\mathbf{\check{n}}$ entirely resides in the intersection of the Bloch sphere and a plane, which is either empty or a point or a circle. Since $\mathcal{C}_{BS}$ is nonempty, we see that the projected image of wave function on the Bloch sphere must be a circle or a point (circle with zero radius) under the flat band condition.

Conversely, if $\mathcal{C}_{BS}$ is either a circle or a point, let $\mathbf{v}$ be its center. Define the vector $\mathbf{p}$ by $\mathbf{p}=\mathbf{v}/(2|\mathbf{v}|^2)$. Then \eq{eq.plane} describes the plane normal to $\mathbf{v}$ with distance from the origin $1/(2|\mathbf{p}|)=|\mathbf{v}|$, so that $\mathcal{C}_{BS}$ lies in this plane. Now, the quadratic form $Q_1,Q_2,Q_3$ in the Hamiltonian clearly satisfy \eq{eq.basis_expansion}, showing that $Q$ is a quadratic form.

We also show that these two conditions are identical in the large $a_1$ limit. In order for $\mathcal{C}_{BS}$ to be a circle, the major diameter and the minor diameter need to be equivalent. From \eq{eq:geopara}, the condition where $d_1=d_2$ is equivalent to $\Delta^2-4b_2^2c_3^2=0$, where $\Delta=b_2^2+b_3^2+c_3^2$. Then we see, $b_2=c_3$ and $b_3=0$ give a circular trajectory.

On the other hand, the quadratic form condition in the large $a_1$ limit amounts to $\tilde{E}_1=E_1$ up to the second order, $\mathcal{O}(\delta^2)$. Since $\tilde{E}$ matches $E_1$ in the terms $k^2, k^2\sin{2\theta}, k^2\cos{2\theta}$, the other terms in $E_1$ such as $k^2\sin{4\theta}, k^2\cos{4\theta}$ need to vanish in order that $\tilde{E}_1=E_1$. According to the supplementary material~\ref{app:Bloch_Sphere}, this gives two equations $b_2^2-b_3^2-c_3^2=0, b_2b_3=0$, which are equivalent to $b_2=c_3, b_3=0$. Thus, the equivalence between the flat band condition and the circular locus condition for the large $a_1$ limit is shown.

\section{Landau Level and Geometric Parameters} \label{sec:LL}
\begin{figure*}[t]
    \centering
    \includegraphics[width=\textwidth]{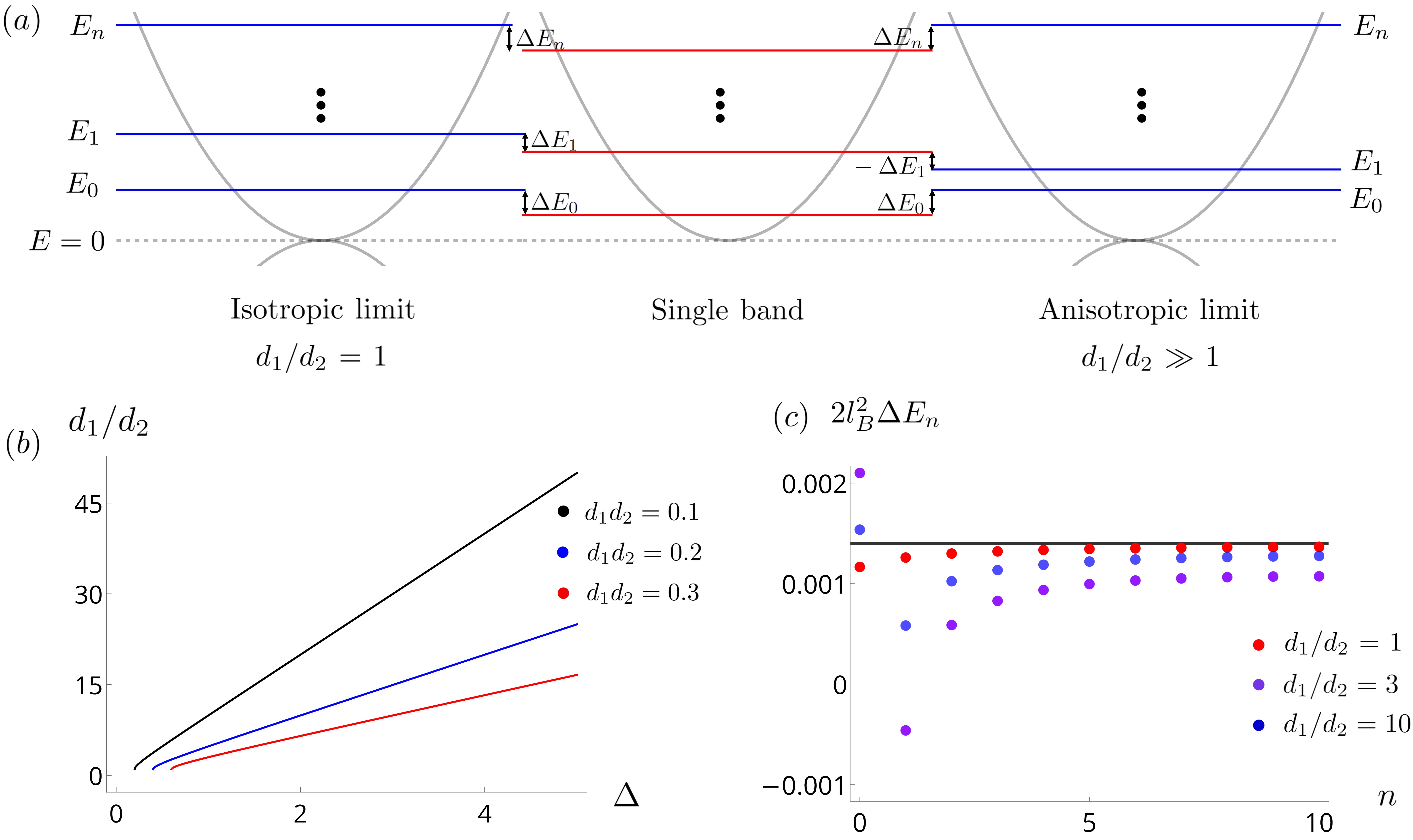}
    \caption{
	(a) Comparison of LL spectra between two-band models (left and right) and a one-band model (middle) in all of which the band with positive curvature has the same mass tensors. When all parameters $a_{1,2,3}, b_{2,3}, c_3$ are positive, $\Delta E_n=E_{n}^{2band}-E_{n}^{1band}\ge0$ when the LL index $n$ is zero or sufficiently large. However, $\Delta E_n$ with small $n$ can be negative if the anisotropy ratio $d_1/d_2$ is large enough. 
	(b) The plot of $d_1/d_2$ as a function of $\Delta=b_2^2+b_3^2+c_3^2.$ 
	Here, $a_1$ is fixed to 1, and $d_1d_2$ is set to 0.1, 0.2, 0.3. $\Delta$ needs to be greater than $4b_2c_3$ since $d_1$ and $d_2$ need to be real numbers in \eq{eq:geopara}. When $\Delta$ is large, $d_1/d_2$ linearly increases with $\Delta$, where the proportionality constant is $(d_1d_2)^{-1}$. (c) $\Delta E_n$ with respect to the LL index, $n$. Three models have different anisotropy factors $d_1/d_2$, but the product of the diameters $d_1d_2$ for all models are fixed to $7.0 \times 10^{-4}$. The black solid line indicates the energy shift corresponding to the $\mathcal{O}(n^0)$ term times $2l_B^2$, which is $2b_2c_3/a_1=1.4\times10^{-3}.$ As the anisotropy factor increases, the energy difference $\Delta E_n$ becomes more negative for initial LLs except for $\Delta E_0$ because of the contribution from the $\mathcal{O}(n^{-1})$ term. As $n$ increases, $\Delta E_n$ converges to the same value for all models.
}
    \label{LandauFig}
\end{figure*}

Thus far, we have presented the definitions of the geometric parameters and a continuum model illustrating the quantum geometry of the wave function in two-band systems with a QBCP. Subsequently, we discuss the physical phenomenon governed by these parameters: the Landau Level problem. Geometric quantities such as the Berry phase are well known to determine the LL, but an extensive relationship between the wave function geometry of a general QBCP Hamiltonian and LL spectra has yet shown. We extend the analysis to the geometric parameters we defined by discussing how the geometric parameters impact the LL spectrum. In particular, we show how the distance parameters, $d_1, d_2$ affect the LL energies. We also analyze how the LL wave function and its quantum geometry change depending on the initial geometry of the QBCP.

\subsection{Landau levels}

To observe the effect of geometrical parameters on the LL, we turn on a magnetic field in the continuum Hamiltonian in \eq{eq:quad_H}. For simplicity, we set $q_i$'s to be zero and assume the energy eigenvalues of the two bands are opposite in sign. To construct the LL Hamiltonian, $H_{LL}$, we substitute the ladder operators in the place of momentum as $k_x \rightarrow (a+a^{\dagger})/(\sqrt{2} l_B), k_y\rightarrow i(a-a^{\dagger})/(\sqrt{2} l_B)$, where $l_B$ is a magnetic length, $l_B=\sqrt{\hbar/eB}$, and $a, a^{\dagger}$ are annihilation and creation operators, which gives
\ba
H_{LL}=&\frac{1}{l_{B}^{2}}(2 a_1 a^{\dagger}a+(a_2+i a_3)a^2+(a_2-i a_3)(a^{\dagger})^{2}+a_1)\sigma_3 \nn \\
&+\frac{1}{l_{B}^{2}}((b_2+i b_3)a^2+(b_2-ib_3)(a^{\dagger})^{2})\sigma_1 \nn \\
&+\frac{1}{l_{B}^{2}}(ic_{3} a^{2}- ic_{3}(a^{\dagger})^{2})\sigma_2.
\label{eq:LandauH}
\ea

We determine the LL spectrum of the continuum model, specifically in the large $a_1$ limit. When all parameters except $a_1$ are zero, the continuum Hamiltonian of \eq{eq:quad_H} becomes a fully decoupled two-band isotropic Hamiltonian, $H(\bk)=a_1k^2\sigma_z$. The LL of this continuum Hamiltonian consists of two distinct Landau levels, denoted by $E_{n,+}$ and $E_{n,-}$, where $E_{n,\pm}$ represents the LL energy of the positive and negative energy bands, respectively. The energy separation between consecutive Landau levels is uniform, and is controlled by the cyclotron frequency $\omega_c$, which is proportional to $a_1$. However, when other parameters besides $a_1$ are present, the LL energy undergoes a shift, which we discuss below.

We evaluate how the geometric parameters affect the LL energies. 
To solely account for the geometric parameters' role, we compare the LL energy of the two-band model with that of a single-band model in a way that the mass tensors of the single band are identical to those of the upper band of the two-band model. 
Since the single band and the upper band of the two-band model have the same energy dispersion, only the geometric parameters or the interband coupling parameters can cause the difference in their LL spectra. 
For a single band Hamiltonian, we use the energy dispersion in \eq{eq:Equadap}, then compare the LL energy of the upper band of the two band model $E^{2band}_{n,+}$ with that of the single-band model $E^{1band}_n$. The details are provided in the \ref{app:LandauCalc}.
The Landau level of the single-band model in \eq{eq:Equadap} is given as
\ba
E^{1band}_n=\hbar \omega_c(n+\frac{1}{2}),
\label{eq:Lsingle}
\ea
where, in the large $a_1$ limit,
\ba
2l_B^2\hbar\omega_c&=4a_1-\frac{2}{a_1}(a_2^2+a_3^2)+\frac{\Delta}{a_1} \nn \\
&=4\alpha-\frac{2}{\alpha}(\beta^2+\gamma^2),
\label{eq:homegac}
\ea
where $\alpha, \beta, \gamma$ are the coefficients in \eq{eq:Equadap}, and $\Delta=b_2^2+b_3^2+c_3^2$.
We note that $a_2, a_3$, and the rest, $b_2, b_3, c_3$, both give the quadratic order correction to the cyclotron frequency, contrary to the order difference in their contribution to the energy dispersion in \eq{eq:tildeE}. The correction from the anisotropic energy dispersion arising from $a_2, a_3$ gives a negative quadratic term, reducing the LL energy spacing.

Now, we compare $E_{n,+}^{2band}$ and $E_n^{1band}$. 
The LLs of the one-band model $E_{n}^{1band}$ is well known to take the form of \eq{eq:Lsingle}, where the spacings between consecutive LLs are identical. However, the spacings between LLs of 2-band model, $E_{n}^{2band}$, are generally not uniform as the LL index $n$ varies. Nevertheless, as $n$ approaches infinity, we will see that the spacings become uniform and converge to $\hbar \omega_c$ of \eq{eq:homegac}. This shows that mass tensors govern the spacings of the LLs. While the effective masses determine the overall spacings of the LLs, interband coupling parameters determine the relative position of the LLs. To show this, we calculate $\Delta E_n=E_{n,+}^{2band}-E_{n}^{1band}$ and observe how this energy difference is expressed in terms of interband coupling parameters. The results are 
\ba
\Delta E_0&=\frac{1}{2l_B^2}(\frac{\Delta}{6a_1}+\frac{4b_2c_3}{3a_1}), \nn \\
\Delta E_1&=\frac{1}{2l_B^2}(-\frac{3\Delta}{10a_1}+\frac{12b_2c_3}{5a_1}), \nn \\
\Delta E_n&=\underbrace{\frac{b_2c_3}{a_1l_B^2}}_{\mathcal{O}(n^{0})}-\underbrace{\frac{1}{2nl_B^2}(\frac{\Delta}{4a_1})}_{\mathcal{O}(n^{-1})}+\mathcal{O}(n^{-2}), (n\ge 2), 
\label{eq:DelE}
\ea
where the large $n$ expansion was performed for $\Delta E_n$. We note that there is no $\mathcal{O}(n)$ term in $\Delta E_n$, confirming that both models' energy spacings are identical. 
$\Delta E_n$ is composed of two terms. 
The first term in $\Delta E_n$ can be rewritten in terms of the geometric parameters $d_1$ and $d_2$ as
\ba
\frac{b_2c_3}{a_1l_B^2}=\frac{a_1d_1d_2}{l_B^2}=-\frac{a_1\Phi_{B}(\mathcal{C}_{BS})}{\pi l_{B}^{2}},
\label{eq:order0term}
\ea
which is proportional to the product of $d_1$ and $d_2$ that is equal to the Berry phase of the BCP when $d_1, d_2 \ll 1$.

Let us compare this result with the semiclassical theory based on Onsager's rule. 
Onsager's rule calculates LLs of dispersive bands by using the formula~\cite{mikitik1999onsager,gao2017onsager,fuchs2018onsager}
\ba
S_0(E_{n})=\frac{2\pi eB}{\hbar}(n+\frac{1}{2}-\frac{\gamma}{2\pi}),
\label{eq:Onsager}
\ea
where $S_0(E_n)$ is the momentum area of the closed curve of energy $E_n$, $e$ is the electron charge, $B$ is the magnetic field, $2\pi\hbar$ is a Planck constant, $n$ is the LL index, and $\gamma$ is the quantum correction including the Berry phase and higher-order responses. The most significant contribution to $\gamma$ comes from the Berry phase of a BCP, giving a constant energy shift to each LL. Our calculation agrees with Onsager's rule, which is confirmed by rewriting the Berry phase in terms of the geometric parameters $d_1, d_2$ as shown in \eq{eq:order0term}. The calculation details are provided in ~\ref{app:Onsager}.

While the product of two distance parameters, $d_1d_2$, affects the constant energy shift, the higher-order correction of $\gamma$ in \eq{eq:Onsager} or the $\mathcal{O}(n^{-1})$ term in \eq{eq:DelE} depends on the ratio of diameters $d_1/d_2$. 
Namely, our theoretical approach expresses the higher-order correction terms by using the geometric parameters.
Explicitly, in \eq{eq:geopara}, we showed that $\Delta$ has a one-to-one correspondence with $d_1/d_2$ when $b_2c_3$ is constant. We illustrate this correspondence in Figure \ref{LandauFig} (b). After fixing $a_1$ and $d_1d_2$, we plot $d_1/d_2$ with respect to $\Delta$. When $\Delta$ is large compared to $b_2c_3$, $\Delta$ is equal to $d_1^2/a_{1}^{2}$. Thus, $\Delta$ and $d_1/d_2$ are proportional to each other because $a_{1}^{2}\Delta/(d_1d_2)=d_1/d_2$. Since the magnitude of the $\mathcal{O}(n^{-1})$ term is proportional to $\Delta$, we observe that the $\mathcal{O}(n^{-1})$ term becomes significant when the ratio $d_1/d_2$ is large. 

The effect of the $\mathcal{O}(n^{-1})$ term is maximized when inspecting the energy levels of initial LLs labeled by small $n$. When the $\mathcal{O}(n^{0})$ term is the dominant contribution to $\Delta E_n$, the difference $\Delta E_n$ is always greater than zero when $b_2c_3\ge0$. On the other hand, when the $\mathcal{O}(n^{-1})$ term is not small, we observe in \eq{eq:DelE} that $\Delta E_n$ can be negative for LL states which satisfy $1\le n \le\frac{\Delta}{8b_2c_3}$, or equivalently $1 \le n\le\frac{d_1}{8d_2}$ Thus, the anisotropy ratio $\frac{d_1}{d_2}$ needs to be greater than $8$ in order for $\Delta E_1$ and other $\Delta E_n$ to have negative values.  This is illustrated in Figure \ref{LandauFig} (c), where the values of $\Delta E_n$ are calculated for several models with different anisotropy factors. The product of $d_1$ and $d_2$ is the same for all models, but their ratios differ, resulting in a difference in initial LL energies. The model with largest anisotropy in Figure \ref{LandauFig} (c) has the anisotropy ratio of $d_1/d_2=10$, causing $\Delta E_1 <0$. 

\subsection{Quantum geometric tensors of the LLs}
We also calculate the quantum geometric tensor of the LL wavefunctions for the two band model. The quantum geometric tensor, $\mathfrak{G}_{ij}$, is a tensor representing the geometric structure of Bloch wavefunctions. The quantum geometric tensor for one band is defined as,
\ba
\mf{G}_{ij}(\bk)
= \brk{\der_i \psi(\bk)}{\der_j \psi(\bk)} - A_i(\bk) A_j(\bk),
\label{eq:qgt_def}
\ea
where $A_i(\bk)=i \brk{\psi(\bk)}{\der_i \psi(\bk)}$ is the Berry connection. The real part of the quantum geometric tensor yields the quantum metric, $\Re[\mathfrak{G}_{ij}]=g_{ij}$,  which provides the information about the infinitesimal distance of quantum states. The imaginary part gives the Berry curvature, $-2\Im[\mathfrak{G}_{ij}]=F_{ij}$. Because the LL Hamiltonian has a magnetic translation symmetry, we can define a Bloch function in the magnetic unit cell that satisfies $|B|a_xa_y=2\pi$. Thus, we can calculate the quantum geometric tensor for the Bloch function. The quantum geometric tensor for the isotropic LL problem has been calculated ~\cite{ozawa2021metriclandau} and we generalize this result to a two-band model with interband coupling, which can be summarized as
\ba
g_{n, xx}(\bk)&=g_{n, yy}(\bk)=(n+\frac{1}{2}+\frac{b_2c_3}{2a_1^2}+\mathcal{O}(n^{-1}))l_B^2, \nn \\
g_{n, xy}(\bk)&=0, \quad F_{n, xy}(\bk)=-l_B^2. (n\ge 2)
\ea
We observe that $\mathfrak{G}_{ij}(\bk)$ is independent of $\bk$. For the entire Landau bands, the Berry curvature is fixed to $-l_B^2$, resulting in a Chern number of $-1$ for each band. Only the diagonal components of the quantum metric change depending on the LL index and the parameter $b_2, c_3$. An additional factor $b_2c_3/2a_1^2$ appearing in the metric is equal to the product $d_1d_2$ in the large $a_1$ limit, which is proportional to the Berry phase.

\begin{table*}
	\centering
	\begin{tabular}{p{0.07\linewidth}p{0.07\linewidth}p{0.2\linewidth}p{0.1\linewidth}p{0.1\linewidth}p{0.13\linewidth}}
		\hline
		Rot Sym. & $|a-b|$ & $h_+(\bk)$ & Geometry & $\mathcal{T}=\mathcal{K}$ &$\mathcal{T}=\sigma_1\mathcal{K}$\\
		\hline
		\hline
		$C_2$ & 0 & $C_{00}+C_{20}k_{+}^{2}+C_{02}k_{-}^{2}$ & Elliptical & Arc & Circle ($d=1$)\\
		& 1 & $C_{10}k_{+}+C_{01}k_{-}$&-&-&-\\
		\hline
		$C_3$ & 0 & $C_{00}+C_{30}k_{+}^{3}+C_{03}k_{-}^{3}$ & Point & Point & - \\
		& 1 & $C_{10}k_{+}+C_{02}k_{+}^2$&Circle&-&Circle ($d=1$)\\
		\hline
		$C_4$ & 0 & $C_{00}$ & Point & Point & -\\
		& 1 & $C_{10}k_+$&-&-&-\\
		& 2 & $C_{20}k_+^2+C_{02}k_-^2$&Elliptical&Arc&Circle ($d=1$)\\
		\hline
		$C_6$ & 0 & $C_{00}$ & Point & Point & -\\
		& 1 & $C_{10}k_+$&-&-&-\\
		& 2 & $C_{20}k_+^2$&Circle&-&Circle ($d=1$)\\
		& 3 & $C_{30}k_+^3+C_{03}k_-^3$&-&-&-\\
		\hline
	\end{tabular}
	\caption{A table showing different shapes of the $\mathcal{C}_{BS}$ depending on the symmetry. For $h_+(\bk)$, the terms above cubic orders are removed. - mark indicates the cases where the time reversal symmetry of a specified form does not commute with the rotation symmetry or BCPs which cannot be quadratic under given symmetry. Circle ($d=1$) indicates that the $\mathcal{C}_{BS}$ becomes a great circle on the Bloch sphere, where the diameter is $1$. $\mathcal{C}_{BS}$ can become an elliptical shape only without time-reversal symmetry.}
	\label{tab:sym}
\end{table*}

\section{Role of Symmetries on quantum geometry and minimal continuum model} \label{sec:Sym}

As shown above, the wave function geometry of the two-band Hamiltonian with a QBCP generally appears in the form of an elliptic curve on the Bloch sphere whose precise form is constrained by the symmetry of the system.
In the following section, we discuss the role of symmetries on the wave function geometry and construct a continuum model that exhibits various quantum geometries of QBCPs. 
Specifically, we review and further extend the results of the previous works~\cite{fang2012qbcp, jung2021flatgeo}, which discuss the form of QBCPs in the presence of symmetries, and show how the wave function geometry of the QBCP Hamiltonian changes under rotation and time-reversal symmetries. Based on this, we construct a QBCP model with $C_4$ symmetry and calculate the Landau levels for different parameters.

First, we consider the rotation symmetry for spinless systems. For the $n$-fold rotation $C_n$ symmetry, the eigenvalues for two bands can be represented by $\lambda_1, \lambda_2=e^{i \frac{2\pi a}{n}},e^{i \frac{2\pi b}{n}}$, respectively, where $a$ and $b$ $\in[0,n-1]$ are integers. The symmetry representation can be connected by unitary transformations or a $U(1)$ phase unless $|a-b|$ is different. Therefore, we can classify $C_n$ symmetries depending on the values of $|a-b|$.

Let us rewrite the Hamiltonian in the $C_n$ eigenbasis, in order to observe how the rotation symmetry enforces the form of the Hamiltonian. We rewrite the Hamiltonian in the following form.
\begin{gather}
	H(\bk)=\sum_{a=0,3,\pm}h_a(\bk)\sigma_a,
\end{gather}
where $\sigma_{\pm}=\frac{1}{2}(\sigma_1 \pm i\sigma_2)$ and $k_{\pm}=k_x \pm ik_y$. To have the $C_n$ symmetry, the Hamiltonian should satisfy $C_n H(\bk)C_n^{-1}=H(C_n \bk)$. For convenience, $h_a(\bk)$ can be expanded in terms of $k_+, k_-$, such that,
\begin{gather}
	h_0(\bk)=\sum A_{ij}k_+^ik_-^j, h_3(\bk)=\sum B_{ij}k_+^ik_-^j, \nn \\
	h_+(\bk)=\sum C_{ij}k_+^ik_-^j,
\end{gather}
and $h_-(\bk)$ is given as complex conjugate of $h_+(\bk)$.
If the rotation symmetry exists, only few terms survive. After expanding in terms of $k_+, k_-$, we write the coefficients that could be nonzero in the presence of $C_n$ symmetry as follows
\begin{gather}
	h_0(\bk)=A_{00}+A_{11}k^2, h_3(\bk)=B_{00}+B_{11}k^2, \nn \\
	h_+(\bk)=\sum_{i-j-a+b\in \mathbb{Z}} C_{ij}k_{+}^{i}k_{-}^{j}.
\end{gather}

We summarize the result in the Table \ref{tab:sym}. For the case where $A_{00}=B_{00}=0$, where the band gap is zero, the geometry of $\mathcal{C}_{BS}$ can either be a point or an elliptic curve or a circle or an arc.  If the wave functions of the BCP become a point on the Bloch sphere, as we mentioned, the wave functions around the BCP converge to a single wave function, which implies that the BCP is non-singular (the band gap could be opened without changing the local wave functions around the BCP by adding a mass term). For other shapes, there is discontinuity at the BCP, prohibiting the convergence of wave functions around the BCP to a point. We illustrate the examples of singular BCPs on the Table \ref{tab:sym}, where $C_2$ and $C_4$ symmetry create an elliptical curve on the Bloch sphere, whereas $C_6$ symmetry create a circular BCP, similar to the flat band condition.

If an additional time reversal symmetry is introduced, more terms vanish, and the geometry is further simplified. For example, in spinless systems, time-reversal symmetry satisfies $\mathcal{T}^2=1$, commuting with a rotation symmetry. Thus, in the basis where the representation of $C_n$ is diagonal, $\mathcal{T}$ is expressed as $\mathcal{K}$ or $\sigma_1\mathcal{K}$. 
When $\mathcal{T}=\mathcal{K}$, we observe a new kind of geometric structure where $\mathcal{C}_{BS}$ becomes an arc. This is an extreme case, where the major diameter $d_1$, is nonzero, but the minor diameter $d_2$ is zero, creating a zero Berry phase around the BCP. The origin of this phenomenon lies on the $\mathcal{T}$ symmetry, where $\mathcal{T}=\mathcal{K}$ enforces $\sigma_2$ coefficients to remain zero, fixing the $\mathcal{C}_{BS}$ on the $xz$-plane. In addition to this condition, since $h_3(\bk)\geq 0$, $\mathcal{C}_{BS}$ remains only on the $z>0$ sector, becoming a semicircle. Therefore $\mathcal{C}_{BS}$ becomes an arc with the combination of $C_2$ and $\mathcal{T}$ symmetry. 

For $\mathcal{T}=\sigma_1\mathcal{K}$, the geometry always becomes the great circle with a diameter $d_1=d_2=1$. In the presence of this symmetry, $h_3(\bk)=0$ and $C_{BS}$ lies on the $xy$-plane with no restriction on $h_{+}(\bk)$. Therefore, $h_{+}(\bk)$ fills all the points on the $x^2+y^2=\frac{1}{4}, z=0$ curve giving a great circle. As shown in the Table \ref{tab:sym}, various geometric shapes can be observed for the case of $C_4$ symmetry with $|a-b|=2$. 

\subsection{Minimal Continuum Model}

We construct a continuum model with minimal parameters which demonstrate various wave function geometry around the band crossing point. As shown in Table~\ref{tab:sym}, a system having $C_4$ symmetry with $|a-b|=2$ exhibits a rich possibility of wave function geometry. If we set the coefficients $B_{11}=a, C_{20}=\frac{1}{2}(b-c), C_{02}=\frac{1}{2}(b+c)$ while others being zero, the corresponding continuum Hamiltonian is given by
\begin{align}	H(\bk)&=a(k_x^2+k_y^2)\sigma_3+b(k_x^2-k_y^2)\sigma_1+c(2k_xk_y)\sigma_2,\nn \\
	&\equiv f_3\sigma_3+f_1\sigma_1+f_2\sigma_2,
	\label{eq:modelconti}
\end{align}
which has the same form as \eq{eq:quad_H} with $a_1=a, b_2=b, c_3=c$, while all the other parameters vanishing.
The corresponding projected trajectory on the Bloch sphere is described by the equation
\begin{gather} 
 \frac{f_1^2}{b^2}+\frac{f_2^2}{c^2}=\frac{f_3^2}{a^2},
	\label{eq:tbgeometery}
\end{gather}
where $f_1, f_2, f_3$ are the coefficients of the Pauli matrices in \eq{eq:modelconti} and the geometric parameters are given by $\{d_1, d_2\}= \{\frac{b}{\sqrt{a^2+b^2}}, \frac{c}{\sqrt{a^2+c^2}}\}$ and $\phi=0$ or $\frac{\pi}{2}$ depending on the magnitude of $b$ and $c$. We note that the curve becomes a circle under the condition $b=c$ at which the continuum Hamiltonian satisfies the quadratic form condition. 

\begin{figure}[t]
    \centering
    \includegraphics[width=0.5\textwidth]{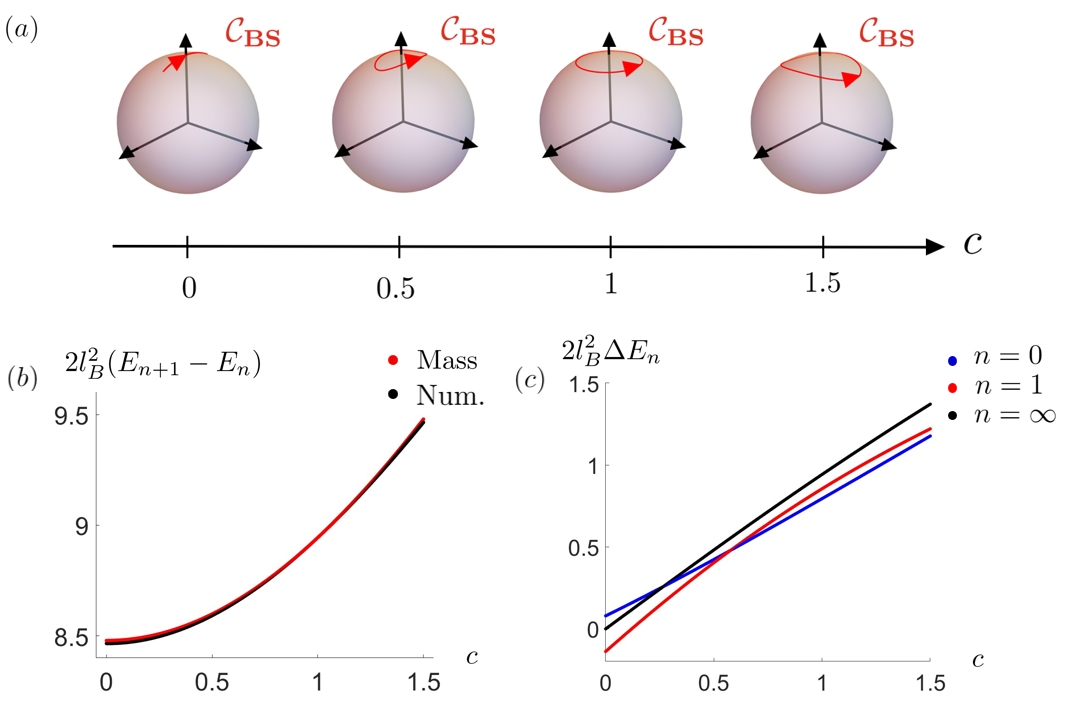}
    \caption{Figures illustrating Landau levels of the upper band described by the minimal continuum Hamiltonian in \eq{eq:modelconti}. $c$ is varied from $0$ to $1.5$ while the other parameters are fixed as $a=2, c=1$. 
    (a) The change in the shape of $\mathcal{C}_{BS}$ as a function of $c$. Here, $d_1$ is fixed to $\frac{1}{\sqrt{5}}$ while $d_2$ increases as $c$ increases. (b) Energy spacings of the LLs. The red line is the spacing calculated from the inverse mass tensor $\alpha$ and the black line is the result from numerical calculation. We use LL with the index $n=100$ to describe the case with sufficiently large $n$. (c) Interband coupling corrections $\Delta E_n = E_n - E_n^{single}$ of the LLs. The correction terms increase with $c$.}
    \label{Fig:LLconti1}
\end{figure}

We inspect the change of the LLs with respect to the anisotropy strength by varying the model parameters. First, we observe how the LLs of dispersive bands described by \eq{eq:modelconti} change upon varying a single parameter that controls the shape of the elliptic trajectory. To this end, we fix $a=2,b=1$ and change $c$ from $0$ to $1.5$. Since the energy of the upper and the lower bands only differ by a sign, it is sufficient to calculate only the LLs $E_n$ of the upper band. By tuning $c$, we change both the mass tensors and the interband coupling parameters. Figure \ref{Fig:LLconti1} (a) illustrates how $\mathcal{C}_{BS}$ is deformed due to the change of $c$, running from an arc ($c=0$) to a circle ($c=1$) via elliptic curves inbetween ($0<c<1$), and then to other elliptic curves elongated along different directions ($1<c<1.5$).

In Section \ref{sec:LL}, we discussed the role of the effective masses and the geometrical parameters in the LL problem. When the LL index $n$ is large, the spacing of the consecutive LLs becomes uniform, and the spacing is solely expressed by using the inverse mass tensors $\alpha, \beta, \gamma$. On the other hand, the interband coupling parameters can shift the relative position of individual LLs, which can be observed by inspecting both small and large $n$ LLs. Thus, by calculating the spacing and relative shift of LLs of a minimal model, we analyze the interplay of the mass tensors and interband coupling parameters.

Before taking the interband coupling into account, let us numerically confirm that the spacings between LLs with large LL indices converge to a constant value determined by the mass tensors as anticipated in Section \ref{sec:QBCP}. To this end, recall that the mass tensors are determined by finding the quadratic form $\alpha(k_x^2+k_y^2)+\beta(k_x^2-k_y^2)+\gamma(2k_xk_y)$ that best approximates a given energy dispersion. For the simple model in \eq{eq:modelconti}, the coefficients $\beta$ and $\gamma$ of the quadratic form vanish. Thus the energy dispersion is best approximated by the isotropic dispersion $\alpha(k_x^2+k_y^2)$, and the spacings between the consecutive LLs are $\hbar\omega_c=\frac{2\alpha}{l_B^2}$. 
In Figure \ref{Fig:LLconti1} (b), we plot the numerical value of LL spacing $(E_{n+1}-E_{n})$ of the upper band at $n=100$ (black) as a function of $c$, and compare it with the theoretical value $\frac{2\alpha}{l_B^2}$ (red). We observe that the two curves match very well. As the inverse effective mass $\alpha(=\frac{2}{m_{xx}}=\frac{2}{m_{yy}})$ increases, the spacing becomes larger as well. Two curves match exactly when the trajectory is circular at $c=1$, that is, when the energy dispersion is exactly quadratic. Note that in this case, isotropic dispersion $E=\alpha(k_x^2+k_y^2)$ is not merely an approximation but an exact result.

Now we turn to the correction due to interband coupling. We compute the difference of LLs between the minimal 2-band model and a single-band model. The LLs of a single-band with isotropic quadratic dispersion is equal to $E^{single}_n=\hbar\omega_c(n+\frac{1}{2})$ as in \eq{eq:Lsingle}. In contrast, the LL spectra of the upper band of the 2-band model do not exactly follow this form since the interband coupling modifies the relative positions of LLs. We calculate $\Delta E_n=E_n-E_n^{single}$ as a function of the parameter $c$, setting the asymptotic spacing between LLs of both models
\begin{gather}
    \lim_{n\to\infty} ( E_{n+1}-E_n ) = \lim_{n\to\infty} ( E^{single}_{n+1}-E^{single}_n ) = \frac{2\alpha}{l_B^2}
\end{gather}
to be identical so that the effect of the interband coupling becomes vivid. In Figure \ref{Fig:LLconti1} (c), we calculate $\Delta E_n$ for various LL indices $n$ with different anisotropy (different $c$ values). First, we observe that $\Delta E_0, \Delta E_{\infty}\ge0$. The qualitative feature of $\Delta E_{\infty}$, represented by the black line in Fig.~\ref{Fig:LLconti1} (c), allows simple explanation: It is proportional to the Berry phase (see \eq{eq:DelE}, \eq{eq:order0term}). When $c=0$, $\mathcal{C}_{BS}$ becomes an arc whose enclosed area is zero. Hence, $\Phi_{B}(\mathcal{C}_{BS})$ grows from zero to a positive value as $c$ increases and $\mathcal{C}_{BS}$ encloses larger area.

In contrast, $\Delta E_1$ is negative for models with a small $c$ as represented by the red line in Fig.~\ref{Fig:LLconti1} (c). As $c$ approaches $0$, the shape of $\mathcal{C}_{BS}$ becomes more anisotropic and the magnitude of $\Delta E_1$ grows in the negative direction. This characteristic was also observed in the large $a_1$ limit. As shown in \eq{eq:DelE}, the parameter $\Delta=b_2^2+b_3^2+c_3^2$, which increases as the $\mathcal{C}_{BS}$ gets more anisotropic, gives a negative contribution to $\Delta E_n$. This observation remains valid even for this minimal model, which lies outside of the large $a_1$ limit. 
However, when $c=1.5$, $\Delta E_1$ takes positive value despite the presence of the anisotropic shape of the elliptic curve. This is due to the large positive contribution to $\Delta E_1$ from the Berry phase. In a general case, the relative position of $E_1$ is determined as a result of the competition between these two contributions: $\Delta E_1$ tends to be negative as the Berry phase is small and the system exhibits a large anisotropy. Since the Berry phase at $c=1.5$ is relatively large, $\Delta E_n$ remains positive despite the presence of the large anisotropy.

\subsection{Nearly flat band}

Next, we calculate the LL spectrum for a nearly flat band coupled to a dispersive band. To construct a model, we use the previous minimal continuum Hamiltonian with parameters $a=2, b=1$ and varying $c$, but add a corresponding $\sigma_0$ term to \eq{eq:modelconti} to flatten the lower band. As discussed in Section \ref{sec:Flat}, when the parameters satisfy the quadratic form condition, it is possible to find an appropriate inverse effective mass $\alpha(a,b,c)$ as a function of other parameters, such that adding $\alpha(k_x^2+k_y^2)\sigma_0$ to \eq{eq:modelconti} creates an exactly flat lower band. For the present continuum model, the quadratic form condition is satisfied when $b=c$ and then adding $\alpha=\sqrt{a^2+b^2}$ term creates an exactly flat band. If the quadratic form condition is not satisfied, we can only create a nearly flat band in a way that the quadratic dispersion from $\sigma_{x,y,z}$ dependent terms is cancelled by $\sigma_0$ dependent term, leading to higher order dispersions and a saddle point at $k=0$. By adding an isotropic quadratic term that approximates the upper band's dispersion, we flatten the lower band and create a nearly flat band.

\begin{figure}[t]
    \centering
    \includegraphics[width=0.5\textwidth]{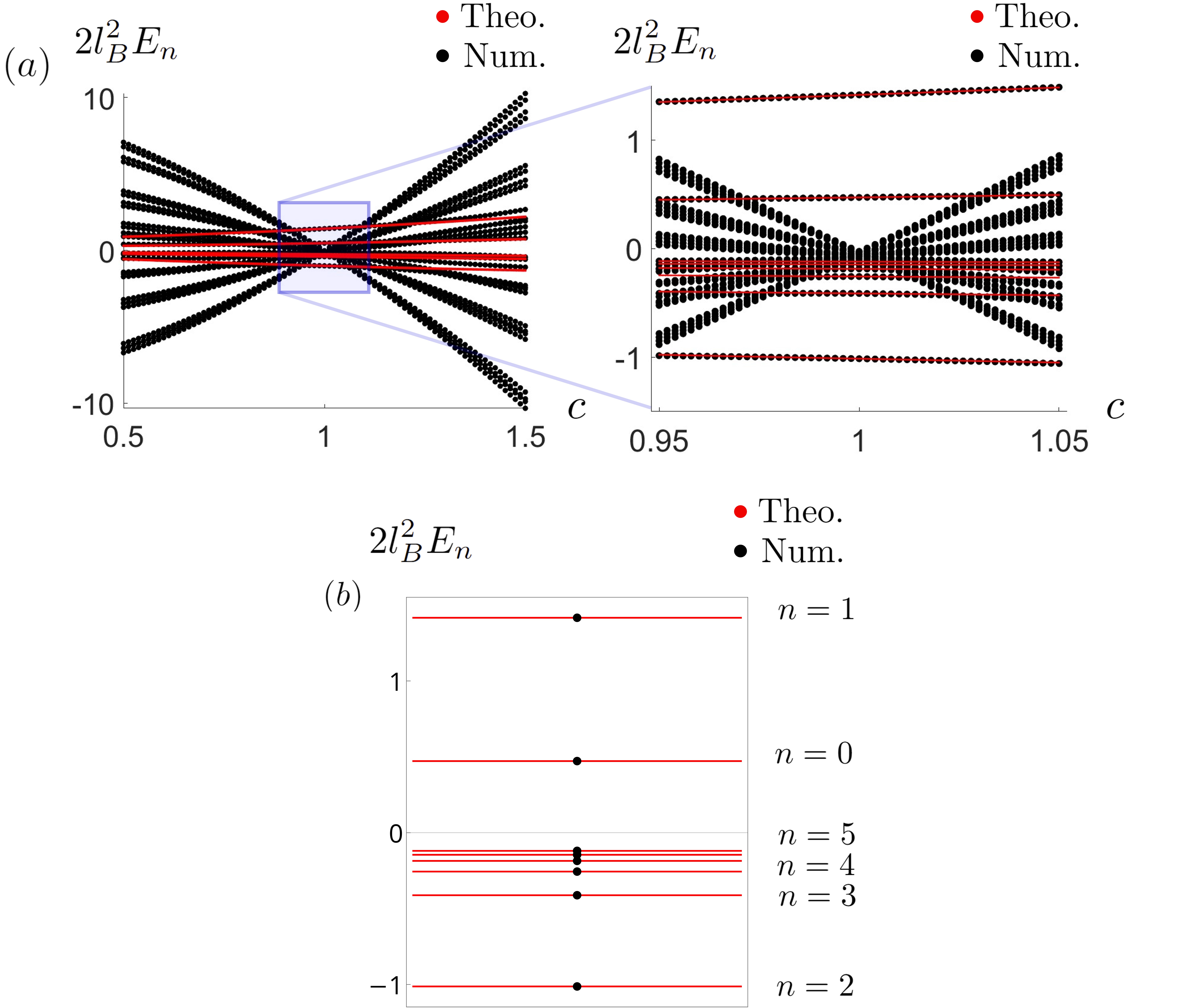}
    \caption{
    	(a) LL spectra of the nearly flat bands. The parameters are $a=2, b=1$, while $c$ varies from $0.5$ to $1.5$ where the flat band limit is satisfied at $c=1$. On the right, we show a magnified figure of the left, where $c$ varies from $0.95$ to $1.05$. The numerical calculation of the LL spectra, represented by black circles, is composed of the LLs created by the interband coupling and those from the energy dispersion. The LLs from the interband coupling obtained by \eq{eq:LLint} are displayed as red lines which match the numerical results well when $c=0.95\sim 1.05$, and when $b$ and $c$ are similiar. (b) LL spectra of the exact flat band when $c=1$. The black dots are the numerical results while the red lines are the analytic results in \eq{eq:LLint}. The first two $E_n$ are positive, while the others are negative. $E_n$ shows a $n^{-1}$ dependence on $n$ when $n\gg 2$.}
    \label{Fig:LLconti2}
\end{figure}

For the exact flat band at $c=1$, the characteristic LL spectrum is illustrated in Figure \ref{Fig:LLconti2} (b). When the flat band condition is satisfied, two LLs from the flat band are positioned above zero energy while the others are below zero energy. The negative LL states are distributed such that the levels are inversely proportional to the Landau level index as shown in Ref.~\cite{rhim2020quantum}. 
We note that the two positive energy LLs are not from the upper band, but they arise mainly from the lower flat band. This LL structure is a characteristic of a flat band coupled to a dispersive band, which was discussed in the case when $d_1=d_2=1$ or $d_1=d_2\ll 1$ in Ref.~\cite{rhim2020quantum}. 

The LL spectrum of the nearly flat band can be considered as a superposition of two different types of LL spectra. One is the characteristic LL spectrum that arises from the interband coupling, and the other is the LL spectrum that comes from the energy dispersion of the nearly flat band. Since the energy dispersion of the lower band is not precisely flat when $c\neq1$, the higher order terms create an equal energy contour which allows us to calculate LL states by using Onsager's rule where the specific levels depend on the energy dispersion over the Brillouin zone ~\cite{roth1966onsager, alex2018saddle}. 

Let us first calculate the LL spectra created by the interband coupling for the nearly flat band. Previous studies only calculated the spectra for the exactly flat band when the interband coupling parameters are either $d_1=d_2=1$ or $d_1=d_2 \ll 1$. We extend this work and provide a general formula that holds for exact flat bands with arbitrary interband coupling parameter $d_1=d_2$ and nearly flat bands with $d_1 \sim d_2$. When the parameter $|b-c|$ of the minimal model is small in magnitude compared to $a, b, c$, the leading terms of the LLs are given
\ba
E_n=&
\begin{cases}
    \frac{1}{l_B^2}(q_1-a), & n=0, \\
    \frac{3}{l_B^2}(q_1-a), & n=1, \\
    \frac{1}{4l_B^2} [\{ (8n-4)\alpha-8a \} \\ 
    -\sqrt{ \{ (8n-4)\alpha-8a \}^2 + 48bc }], & n \ge 2 \\
\end{cases} \nn \\
\label{eq:LLint}
\ea
where $n=0, 1$ represent LLs of the nearly flat bands which are located above the zero energy and $n \ge 2$ express LLs that are below the zero energy. 
This equation holds when $b$ and $c$ are similar or equivalently when $d_1 \sim d_2$. Figure \ref{Fig:LLconti2} (a) shows the numerical results of the LL spectra and the analytic results in \eq{eq:LLint} for $c\in[0.5, 1.5]$. The two results match well when $b$ and $c$ are similar. We note that LLs created by the interband coupling exist not only at exactly flat band limit ($c=1$) but also for nearly flat band limit. The spacings between these levels increase with $c$, $d_1$, and $d_2$. 
To show the dependence of the LLs on the interband coupling parameters $d_1, d_2$, we expand \eq{eq:LLint} in the large $n$ limit and express $b, c$ in terms of $d_1, d_2$, which leads to
\ba
E_n=-\frac{3a}{4nl_B^2}\frac{d_1d_2}{\sqrt{1-\frac{d_1^2}{2}-\frac{d_2^2}{2}}} \quad (n\ge2),
\ea
where $E_n$ is inversely proportional to the LL index $n$. In the nearly flat band, the LL spectra formed by the interband coupling also show $n^{-1}$ dependence on the LL index, which generalizes the result of the flat band limit shown in Ref. \cite{rhim2020quantum}.

In addition to the LLs created from the interband coupling, we observe the LLs created from the energy dispersion of the nearly flat bands, which scale linearly with $|c-1|$. By expanding the dispersion of the nearly flat band in $\mathcal{O}(c-1)$, we find that the leading term $-\frac{(c-1)}{4\sqrt{5}}\frac{k_x^4+k_y^4}{k_x^2+k_y^2}$ has a linear dependence on $c-1$. Thus, when the band is nearly flat, or when $b$ and $c$ are close in value, the energy dispersion is proportional to $c-1$. 

\section{Conclusion}

In this work, we have studied the interplay between the interband coupling and the wave function geometry in two-band Hamiltonians with QBCPs. Among the nine parameters that exist in a generic two-band quadratic Hamiltonian with a QBCP, six corresponds to the mass tensors of the two bands, while the remaining three describe the interband coupling. More explicitly, in terms of the elliptic trajectory on the Bloch sphere corresponding to the collection of the wave functions in momentum space, we found that the three interband coupling parameters determine the major $d_1$ and minor $d_2$ diameters of the elliptic curve and the canting angle $\phi$ of the ellipse with respect to the Bloch sphere.
Depending on the presence or absence of the singularity at the QBCP, the elliptic trajectory takes different shapes, including point, arc, circle, ellipse, etc. In particular, the quadratic form condition (or flat band condition) simplifies the geometry of the wave function and gives only a circular form of the trajectory on the Bloch sphere.

In addition, we explored the influence of interband couplings on the LL spectrum. By comparing two systems that share the same mass tensors with distinct interband couplings, we demonstrated that the interband couplings induce a constant energy shift of LLs as well as the energy of initial LLs near the QBCP. Moreover, we calculated the quantum geometric tensors of the LL wave functions and examined the role of the interband coupling.

We also extended our analysis to lattice models and examined the influence of rotation and time-reversal symmetries on the wave function geometry. We designed a minimal continuum model with $C_4$ rotation symmetry which showcases various geometric structures of wave functions. We calculate the LL spectra of the model and show how interband coupling parameters modify the LLs for dispersive bands and create unique LL structure for nearly flat bands.

Our work clearly demonstrates that the quantum geometry of the wave function contains the essential information of the complicate interband coupling effect. Revealing the novel physical responses, beyond the LL problem in QBCPs, induced by nontrivial quantum geometry are definitely important issues for further investigation, which we leave for future studies.

\magenta{Acknowledgement}
J.J., H.L., and B.J.Y. were supported by the Institute for Basic Science in Korea (Grant No. IBS-R009-D1),
Samsung  Science and Technology Foundation under Project Number SSTF-BA2002-06,
the National Research Foundation of Korea (NRF) grant funded by the Korean government (MSIT) (No.2021R1A2C4002773, and No. NRF-2021R1A5A1032996). 


\let\oldaddcontentsline\addcontentsline
\renewcommand{\addcontentsline}[3]{}

\bibliography{Refs}

\let\addcontentsline\oldaddcontentsline

\clearpage
\onecolumngrid
\begin{center}
\textbf{\large Supplemental Material for ``\ourtitle"}
\end{center}
\setcounter{section}{0}
\setcounter{figure}{0}
\setcounter{equation}{0}
\renewcommand{\thefigure}{S\arabic{figure}}
\renewcommand{\theequation}{S\arabic{equation}}
\renewcommand{\thesection}{S\arabic{section}}

\section{Quantum geometry and the Bloch sphere \label{app:Bloch_Sphere}}
For general QBCP in a form of \eq{eq:Equad}, the energy dispersion $E_-^2(\bk)$ is
\ba
E_{-}^{2}(\bk)=&k^4(a_1^2+\frac{a_2^2+a_3^2+b_2^2+b_3^2+c_3^2}{2}+2a_1a_2\cos{2\theta} \nn \\ &+2a_1a_3\sin{2\theta}+\frac{a_2^2+b_2^2-a_3^2-b_3^2-c_3^2}{2}\cos{4\theta} \nn \\
&+(a_2a_3+b_2b_3)\sin{4\theta}),
\ea
where $k^2=k_x^2+k_y^2$ and $\theta=\arctan{k_y/k_x}$.
In the large $a_1$ limit, $E_-$ can be expanded in terms of $\delta$, where $\delta$ is the ratio between other coefficients and $a_1$,
\ba
E_-=&a_1k^2(1+\frac{2a_2}{a_1}\cos{2\theta}+\frac{2a_3}{a_1}\sin{2\theta}+\frac{a_2^2+a_3^2+b_2^2+b_3^2+c_3^2}{2a_1^2}+\frac{a_2^2+b_2^2-a_3^2-b_3^2-c_3^2}{2a_1^2}\cos{4\theta}+\frac{a_2a_3+b_2b_3}{a_1^2}\sin{4\theta})^\frac{1}{2} \nn \\
=&a_1k^2(1+\frac{1}{2}(\frac{2a_2}{a_1}\cos{2\theta}+\frac{2a_3}{a_1}\sin{2\theta}+\frac{a_2^2+a_3^2+b_2^2+b_3^2+c_3^2}{2a_1^2}+\frac{a_2^2+b_2^2-a_3^2-b_3^2-c_3^2}{2a_1^2}\cos{4\theta}+\frac{a_2a_3+b_2b_3}{a_1^2}\sin{4\theta}) \nn \\
&-\frac{1}{8} (\frac{2a_2}{a_1}\cos{2\theta}+\frac{2a_3}{a_1}\sin{2\theta}+\frac{a_2^2+a_3^2+b_2^2+b_3^2+c_3^2}{2a_1^2}+\frac{a_2^2+b_2^2-a_3^2-b_3^2-c_3^2}{2a_1^2}\cos{4\theta}+\frac{a_2a_3+b_2b_3}{a_1^2}\sin{4\theta})^2 +....) \nn \\
=&a_1k^2(\underbrace{1}_{\mathcal{O}(\delta^{0})}+\underbrace{\frac{a_2}{a_1}\cos{2\theta}+\frac{a_3}{a_1}\sin{2\theta}}_{\mathcal{O}(\delta)}+\underbrace{\frac{b_2^2+b_3^2+c_3^2}{4a_1^2}+\frac{b_2^2-b_3^2-c_3^2}{4a_1^2}\cos{4\theta}+\frac{b_2b_3}{2a_1^2}\sin{4\theta}}_{\mathcal{O}(\delta^2)}+...).
\ea
We note that when $E_-$ is expanded in $\delta$, the coefficients up to $\delta^2$ match \eq{eq:tildeE}, except for the $\cos{4\theta}, \sin{4\theta}$ terms. Since $k^2\cos{4\theta}$ and $k^2\sin{4\theta}$ cannot be expressed as a quadratic expression, they are neglected in \eq{eq:tildeE}.

\section{Calculation of the Geometrical Parameters \label{app:GeoCalc}}
We explain how to calculate the geometrical parameters. The three Bloch coefficients $f_1, f_2, f_3$ in \eq{eq:Ham_fvec} of a generic quadratic two-band Hamiltonian, satisfy the quadratic equation $f_iM_{ij}f_j=0$, where the matrix is
\ba
M_{ij}=\begin{pmatrix} \frac{1}{a_1^2} & -\frac{a_2}{a_1^2b_2} & \frac{a_2b_3}{a_1^2b_2c_3}-\frac{a_3}{a_1^2c_3} \\ * & -\frac{1}{b_2^2}+\frac{a_2^2}{a_1^2b_2^2} & \frac{a_2a_3}{a_1^2b_2b_3}+\frac{b_3}{b_2^2c_3}-\frac{a_2^2b_3}{a_1^2b_2^2c_3} \\ * & * &-\frac{1}{c_3^2}(1-\frac{a_3^2}{a_1^2}+\frac{2a_2a_3b_3}{a_1^2b_2}+\frac{b_3^2}{b_2^2}-\frac{a_2^2b_3^2}{a_1^2b_2^2}) \end{pmatrix}, 
\label{eq:Mmatrixo}
\ea
and matrix elements shown as $*$ are elements that can be acquired by the symmetry of the matrix, $M_{ij}=M_{ji}$. We explain how to obtain $M_{ij}$.

Unnormalized Bloch vectors, $f_i$, of \eq{eq:quad_H} is written as
\ba
f_1&=a_1(k_x^2+k_y^2)+a_2(k_x^2-k_y^2)+a_3(2k_xk_y), \nn \\
f_2&=b_2(k_x^2-k_y^2)+b_3(2k_xk_y), \nn \\
f_3&=c_3(2k_xk_y).
\label{eq:Blochvec}
\ea
From \eq{eq:Blochvec}, we can express $k_x^2+k_y^2, k_x^2-k_y^2, 2k_xk_y$ in terms of $f_1, f_2, f_3$ as following
\ba
k_x^2+k_y^2&=\frac{1}{a_1}f_1-\frac{a_2}{a_1b_2}f_2+(\frac{a_2b_3}{a_1b_2c_3}-\frac{a_3}{a_1c_3})f_3, \nn \\
k_x^2-k_y^2&=\frac{1}{b_2}f_2-\frac{b_3}{b_2c_3}f_3, \nn \\
2k_xk_y&=\frac{1}{c_3}f_3.
\ea
Then, \eq{eq:Mmatrixo} is obtained by the identity, $(k_x^2+k_y^2)^2=(k_x^2-k_y^2)^2+(2k_xk_y)^2$.

$\mathcal{C}_{BS}$ is formed by the the intersection of the surface $f_iM_{ij}f_{j}=0$ and $S^2_{BS}$, resulting an elliptic curve. $d_1, d_2, \phi$ can be calculated in the large $a_1$ limit using $M_{ij}$. In this limit, the matrix reduces to
\ba
\tilde{M}_{ij}=\begin{pmatrix} \frac{1}{a_1^2} & 0 & 0 \\ * & -\frac{1}{b_2^2} & \frac{b_3}{b_2^2c_3} \\ * & * &-\frac{1}{c_3^2}(1+\frac{b_3^2}{b_2^2}) \end{pmatrix},
\label{eq:Mmatrix}
\ea
and its eigenvalues and eigenvectors are obtained
\ba
&\lambda_1=\frac{1}{a_1^2}, \lambda_{\pm}=-\frac{\Delta\pm\sqrt{\Delta^2-4b_2^2c_3^2}}{2b_2^2c_3^2}, \nn \\
&v_1=\begin{pmatrix} 1 \\ 0 \\ 0 \end{pmatrix}, 
v_{-}=\mathcal{N}\begin{pmatrix} 0 \\ b_3c_3 \\ c_3^2-\frac{\Delta}{2}+\frac{\sqrt{\Delta^2-4b_2^2c_3^2}}{2} \end{pmatrix},
v_{+}=\mathcal{N}\begin{pmatrix} 0 \\ c_3^2-\frac{\Delta}{2}+\frac{\sqrt{\Delta^2-4b_2^2c_3^2}}{2} \\ -b_3c_3 \end{pmatrix},
\ea
where $\Delta=b_2^2+b_3^2+c_3^2$ , $b_2>c_3$, and $\mathcal{N}$ is a proper normalization constant of $v_-, v_+$. $\phi$ is obtained by $f_2, f_3$ components of $v_2$ giving, 
\ba
&\phi=\arctan{\frac{c_3^2-\frac{\Delta}{2}+\frac{\sqrt{\Delta^2-4b_2^2c_3^2}}{2}}{b_3c_3}} \mod \pi.
\ea
If $b_3=0$, two axes of ellipse align with $f_2, f_3$ axis, resulting in $\phi=0$ or $\frac{pi}{2}$, depending on the magnitude of $b_2$ and $c_3$.
Also, when $b_2$ or $c_3=0$, the corresponding Bloch components $f_2, f_3$ becomes zero, so $\phi$ becomes 0 or $\frac{\pi}{2}$ respectively, which can be calculated without using $M_{ij}$.

To calculate $d_1$ and $d_2$, we rotate $f_1, f_2,$ and $f_3$ by the appropriate orthogonal matrix to obtain two equations. The first equation describes the surface of the Bloch sphere and the second equation provides the equation of the elliptic cone.
\ba
\tilde{f}_1^2+\tilde{f}_2^2+\tilde{f}_3^2=\frac{1}{4}, \nn \\
\lambda_1\tilde{f}_1^2=|\lambda_2|\tilde{f}_2^2+|\lambda_3|\tilde{f}_3^2.
\ea
The curve, $\mathcal{C}_{BS}$, formed by the intersection of two equations form an ellipse when projected to $f_2f_3$-plane. The equation of an ellipse is given as, $\frac{f_2^2}{(d_1/2)^2}+\frac{f_3^2}{(d_2/2)^2}=1$, where $d_1, d_2$ are
\ba
d_1^2=\frac{\lambda_1}{\lambda_1+|\lambda_{-}|}, d_2^2=\frac{\lambda_1}{\lambda_1+|\lambda_{+}|}.
\label{eq:dlambda}
\ea
When $b_3=0$, $d_1=\max(|\frac{b_2}{a_1}|, |\frac{c_3}{a_1}|)$, $d_2=\min(|\frac{b_2}{a_1}|, |\frac{c_3}{a_1}|)$. We have computed $d_1$, $d_2$, and $\phi$ and refer to these quantities as the geometrical parameters or the interband coupling parameters interchangeably.

\section{Construction of a 
 Minimal Continumm Model using Kronecker-product Construction
\label{app:Minmodel}}
We propose a simple continuum quadratic band touching model, which is not in the large $a_1$ limit, where the geometric parameters can be continuously tuned. We design the model using a Kronecker-product construction (KPC) ~\cite{hwang2021flatgeo}, a method creating a flat band Hamiltonian using basis molecular orbitals $\ket{\psi_1}, \ket{\psi_2}$. ($\ket{\psi_1}$ and $\ket{\psi_2}$ do not need to be normalized.) By writing a Hamiltonian in this form,
\begin{gather}
    H_{flat}(k_x,k_y)=\ket{\psi_1}\bra{\psi_1}.
\end{gather}
we can guarantee that $H_{flat}\ket{\psi_2}=0$ regardless of the normalization condition. We set $\ket{\psi_1}$ and $\ket{\psi_2}$ as

\ba
\ket{\psi_1}=\begin{pmatrix} k_x-itk_y \\ k_y \end{pmatrix}, \ket{\psi_2}=\begin{pmatrix} -k_y \\ k_x+itk_y \end{pmatrix},
\ea
where $t$ is a continuous parameter.

Using KPC, we write
\begin{gather}
    H_t(k_x, k_y)
    =\begin{pmatrix} k_x^2+t^2 k_y^2 & k_xk_y-itk_y^2 \\ k_xk_y+itk_y^2 & k_y^2\end{pmatrix},
    \label{eq:minflat}
\end{gather}

where the coefficients are given as

\begin{gather}
    q_1=\frac{1}{2}, q_2=0, q_3=\frac{1}{2}(t^2+1), a_1=\frac{1}{2}, a_2=0, \nn \\ 
    a_3=\frac{1}{2}(t^2-1), b_2=1, b_3=0, c_3=t. \nn \\
    \label{eq:cond}
\end{gather}
We check that the flatness condition ~\cite{rhim2020quantum}, $b_3=\frac{a_2b_2}{2a_1}, c_3=\pm \frac{b_2}{2a_1}\sqrt{4a_1a_3+b_2^2-a_2^2}$,  is indeed satisfied   and the maximal quantum distance $d_{max}$ is given as

\begin{gather}
    d_{max}=\frac{|b_2|}{\sqrt{2b_2^2+4a_1a_3-a_2^2}}=\frac{1}{\sqrt{1+t^2}}.
\end{gather}
Thus, We have a two-band flat band Hamiltonian with tunable $d_{max}$ using parameter $t$ in \eq{eq:minflat}. Next, we construct a minimal continuum Hamiltonian, $H_{t,\beta}$, where the anisotropy factor, $d_1/d_2$, can also be tuned by an additional parameter $\beta$.
\begin{gather}
    H_{t,\beta}(k_x, k_y)=
    (\frac{1}{2}k_x^2+\frac{1}{2}(t^2+1)k_y^2)\sigma_0 \nn \\
    +(\frac{1}{2}k_x^2+\frac{1}{2}(\beta t^2-\frac{1}{\beta})k_y^2)\sigma_x \nn \\
    +k_xk_y \sigma_x+ t k_y^2 \sigma_y.
\end{gather}
Here, the other parameters are equivalent to \eq{eq:cond} except $a_3$. By changing $a_3=\frac{1}{2}(\beta t^2-\frac{1}{\beta})$, we calculate how $\beta$ affects the anisotropy from calculating the characteristic equation of matrix $M_{ij}$. After substituting the parameters, the characteristic equation becomes
\begin{gather}
    f(\lambda)=(\frac{1}{2}t^2-\lambda)(\lambda^2-t_3\lambda-\frac{1}{4}t^2)=0.
\end{gather}
For the parameter set in \eq{eq:cond} with $a_3=\frac{1}{2}(t^2-1)$, the roots of the equation are $\lambda=\frac{1}{2}t^2, \frac{1}{2}t^2, -\frac{1}{2}$.
Since two of the eigenvalues are equal in this case, the projection of the wave function onto the Bloch sphere becomes a circle. When $a_3=\frac{1}{2}(\beta t^2-\frac{1}{\beta})$, the eigenvalues are $\frac{\beta}{2}t^2, \frac{1}{2}t^2, -\frac{1}{2\beta}$, changing the shape of the loop. Since the two eigenvalues with the same sign do not have a same magnitude, the projection is no longer a circle. The two diameter of this curve is controlled by $t,\beta$ so that 
\begin{gather}
    d_1=\sqrt{\frac{1}{1+\beta^2t^2}}, d_2=\sqrt{\frac{1}{1+\beta t^2}},
\end{gather}
and the ratio becomes
\begin{gather}
    \frac{d_1}{d_2}=\sqrt{\frac{1+\beta t^2}{1+ \beta^2t^2}}.
\end{gather}
We see that if $\beta=1$, then $d_1/d_2=1$ and as $\beta$ deviates from 1, the ratio gets deviates from 1 as well.

\section{Landau Level Energy and Quantum Metric
\label{app:LandauCalc}}
Landau level of a continuum Hamiltonian $H(\bk)$,
\ba
H(\bk)=&k^2(a_1+a_2\cos{2\theta}+a_3\sin{2\theta})\sigma_3 \nn \\ 
&+k^2(b_2\cos{2\theta}+b_3\sin{2\theta})\sigma_1+k^2(c_3\sin{2\theta})\sigma_2,
\ea
can be acquired by substituting $k_x=(a+a^{\dagger})/(\sqrt{2}l_B), k_y=i(a-a^{\dagger})/(\sqrt{2}l_B)$. With this substitution, we write the Landau level Hamiltonian $H_{LL}$ as \eq{eq:LandauH}, which is written as
\ba
H_{LL}=\frac{1}{2l_B^2}\begin{pmatrix} h_0 & 0 & g_0 & 0 & \dots \\
0 & h_1 & 0 & g_1 & \dots \\
g_0^{\dagger} & 0 & h_2 & 0 \dots \\
0 & g_1^{\dagger} & 0 & h_3 & \dots \\
\vdots & \vdots & \vdots & \vdots & \ddots \end{pmatrix},
\label{eq:HLLBlock}
\ea
where,
\ba
h^{2band}_n=(4n+2)a_1\sigma_3=(4n+2)\begin{pmatrix} a_1 & 0 \\ 0 & -a_1 \end{pmatrix},
\ea
and
\ba
g^{2band}_n=&2\sqrt{(n+1)(n+2)}((a_2+ia_3)\sigma_3+(b_2+ib_3)\sigma_1 \nn \\
&+(2ic_3)\sigma_2) \nn \\
=&2\sqrt{(n+1)(n+2)}\begin{pmatrix} a_2+ia_3 & b_2+c_3+ib_3 \\ b_2-c_3+ib_3 & -a_2-ia_3 \end{pmatrix}.
\ea
In this basis, a state $\ket{\psi}$, represented by a column vector
\ba
\ket{\psi}=\begin{pmatrix} C_0 \\ D_0 \\ C_1 \\ D_1 \\ \vdots \end{pmatrix}=\sum_{n\ge0,+} C_n \ket{n,+}+\sum_{n\ge0,-} D_n\ket{n,-},
\label{eq:LLwaveftn}
\ea
is composed of a superposition of LL wavefunctions where $\ket{n,+}$(or $\ket{n,-}$) is the upper band(or the lower band) Landau level wavefunctions of an isotropic Hamiltonian, $H(\bk)=a_1k^2\sigma_z$. Let's indicate the $n^{th}$ Landau level with positive energy as $E_{n,+}$ and negative energy as $E_{n,-}$. In the large $a_1$ limit, we calculate the energy levels using perturbation theory. First, the $0^{th}$ order energy corresponds to the diagonal component of $H_{LL}$, so $E^{0}_{n,\pm}=\pm\frac{(4n+2)a_1}{2l_B^2}$. Next order of energy is calculated using second order perturbation theory.
\ba
E^{2band}_{n,+}=&E^0_{n,+}+\Delta E_{n,+}^2=E^0_{n,+}+\sum_{(m,+)} \frac{|H_{nm}|^2}{E^0_{n,+}-E^0_{m,+}}+\sum_{(m,-)} \frac{|H_{nm}|^2}{E^0_{n,+}-E^0_{m,-}} \nn \\
=&\frac{1}{2l_B^2}((4n+2)a_1+\frac{4(a_2^2+a_3^2)(n-1)n}{a_1(4n+2)-a_1(4n-6)}+\frac{4(a_2^2+a_3^2)(n+1)(n+2)}{a_1(4n+2)-a_1(4n+10)} \nn \\
&+\frac{4((b_2-c_3)^2+b_3^2)(n-1)n}{a_1(4n+2)+a_1(4n-6)}+\frac{4((b_2+c_3)^2+b_3^2)(n+1)(n+2)}{a_1(4n+2)+a_1(4n+10)}) \nn \\
=&\underbrace{\frac{n}{2l_B^2}(4a_1-\frac{2}{a_1}(a_2^2+a_3^2)+\frac{\Delta}{a_1})}_{\mathcal{O}(n)}+\underbrace{\frac{1}{2l_B^2}(2a_1-\frac{1}{a_1}(a_2^2+a_3^2)+\frac{\Delta}{2a_1}+\frac{2b_2c_3}{a_1})}_{\mathcal{O}(n^{0})}+\underbrace{\frac{1}{2nl_B^2}(-\frac{\Delta}{4a_1})}_{\mathcal{O}(n^{-1})}+\mathcal{O}(\frac{1}{n^2}),
\label{eq:En2plong}
\ea
\label{eq:Lenergy2band}
where $\Delta=b_2^2+b_3^2+c_3^2$.
When $n\le 1$, \eq{eq:En2plong} does not hold because the shift of the $n^{th}$ LL energy of the upper band only comes from the $(n+2)^{th}$ LL and not from the $(n-2)^{th}$ LL. Since there is no $(n-2)^{th}$ Landau level when $n \le 1$, we calculate $E_{0,+}, E_{1,+}$ separately which gives

\ba
E^{2band}_{0,+}=&\frac{1}{2l_B^2}(2a_1-\frac{a_2^2+a_3^2}{a_1}+\frac{2\Delta}{3a_1}+\frac{4b_2c_3}{3a_1}), \nn \\
E^{2band}_{1,+}=&\frac{1}{2l_B^2}(6a_1-\frac{3(a_2^2+a_3^2)}{a_1}+\frac{6\Delta}{5a_1}+\frac{12b_2c_3}{5a_1}),
\ea
and $E^{2band}_{n,-}=-E^{2band}_{n,+}$ for all $n$.

Now, we compare this energy to the single band continuum model which approximates the energy of the two band, where the Hamiltonian is given as \eq{eq:tildeE}. For this LL Hamiltonian, same block Hamiltonian form from \eq{eq:HLLBlock} is retained while the block matrix is changed.
\ba
h_n=(4n+2)(a_1+\frac{\Delta}{4a_1}), g_n=\sqrt{(n+1)(n+2)}(a_2+ia_3).
\ea

For the single band $H_{LL}$, $h_n$ and $g_n$ are given as a number instead of 2 by 2 matrices. In the large $a_1$ limit, we calculate $E^{1band}_n$.
\ba
E^{1band}_n=&\frac{1}{2l_B^2}((4n+2)(a_1+\frac{\Delta}{4a_1})+\frac{4(a_2^2+4a_3^2)(n-1)n}{a_1(4n+2)-a_1(4n-6)}+\frac{4(a_2^2+a_3^2)(n+1)(n+2)}{a_1(4n+2)-a_1(4n+10)}) \nn \\
=&\underbrace{\frac{n}{2l_B^2}(4a_1-\frac{2}{a_1}(a_2^2+a_3^2)+\frac{\Delta}{a_1})}_{\mathcal{O}(n)}+\underbrace{\frac{1}{2l_B^2}(2a_1-\frac{1}{a_1}(a_2^2+a_3^2)+\frac{\Delta}{2a_1})}_{\mathcal{O}(1)},
\ea
which hold for $n \ge 2$, and when $n \le 1$,
\ba
E^{1band}_0&=\frac{1}{2l_B^2}(2a_1-\frac{a_2^2+a_3^2}{a_1}+\frac{\Delta}{2a_1}), \nn \\
E^{1band}_1&=\frac{1}{2l_B^2}(6a_1-\frac{3(a_2^2+a_3^2)}{a_1}+\frac{3\Delta}{2a_1}).
\ea

We also calculate the quantum geometric tensor of LL wavefunction for the large $a_1$ limit, with $a_2, a_3=0$. To do so, we need to obtain the periodic part of the wave functions for the LL Hamiltonian, which we denote as $\ket{\tilde{u}_{n,\bk}}$. This wave function represent periodic part of the eigenfunction $\ket{n}$, with a momentum $\bk$. In Ref. ~\cite{ozawa2021metriclandau}, the explicit form of the wave function and its inner product with its derivative is calculated and we use the result of the calculation:

\ba
&\langle\tilde{u}_{m,\bk}|\partial_x\tilde{u}_{n,\bk}\rangle=-i\sqrt{\frac{D_n}{D_m}}(\frac{l_B}{2}\delta_{m+1,n}+k_yl_B^2\delta_{m,n}+ml_B\delta_{m-1,n}), \nn \\
&\langle\tilde{u}_{m,\bk}| \partial_y\tilde{u}_{n,\bk}\rangle=\sqrt{\frac{D_m}{D_n}}(\frac{l_B}{2}\delta_{m,n+1}-nl_B\delta_{m,n-1}), \nn \\ 
&\langle\partial_x\tilde{u}_{m,\bk}|\partial_x\tilde{u}_{n,\bk}\rangle=\sqrt{\frac{D_n}{D_m}}(\frac{l_B^2}{4}\delta_{m+2,n}+l_B^3k_y\delta_{m+1,n}+((m+\frac{1}{2})l_B^2+l_B^4k_y^2)\delta_{m,n}+2ml_B^3k_y\delta_{m-1,n}+m(m-1)l_B^2\delta_{m-2,n}), \nn \\
&\langle\partial_x\tilde{u}_{m,\bk}|\partial_y\tilde{u}_{n,\bk}\rangle=i\sqrt{\frac{D_m}{D_n}}(\frac{l_B^2}{4}\delta_{m,n+2}+\frac{l_B^3}{2}k_y\delta_{m,n+1}+\frac{l_B^2}{2}\delta_{m,n}-nl_B^3k_y\delta_{m,n-1}-n(n-1)l_B^2\delta_{m,n-2}), \nn \\
&\langle\partial_y\tilde{u}_{m,\bk}|\partial_y\tilde{u}_{n,\bk}\rangle=\sqrt{\frac{1}{D_nD_m}}((\frac{l_B^2}{4}D_{m+1}+mnl_B^2D_{m-1})\delta_{m,n}-\frac{m}{2}l_B^2D_{m-1}\delta_{m,n+2}-\frac{n}{2}l_B^2D_{m+1}\delta_{m+2,n}),
\label{eq:Landauwav}
\ea
where $D_m=2^mm!\sqrt{\pi}$. Using \eq{eq:Landauwav}, we calculate the quantum geometric tensor for the wave function composed of the superposition of the LL wave function. If the wave function is written as $\ket{\psi}=\sum_{n\ge0}c_{n}\ket{n}$, then the quantum geometric tensor is the following:
\ba
\mathcal{G}_{xx}=&l_B^2\sum_{\alpha}(|c_{\alpha}|^2(\alpha+\frac{1}{2})+(c^{*}_{\alpha+2}c_{\alpha}+c_{\alpha+2}c^*_{\alpha})\frac{\sqrt{(\alpha+2)(\alpha+1)}}{2})
-l_B^2(\sum_{\alpha}(c^{*}_{\alpha+1}c_{\alpha}+c_{\alpha+1}c^*_{\alpha})\sqrt{\frac{(\alpha+1)}{2}})^2, \nn \\
\mathcal{G}_{xy}=&il_B^2\sum_{\alpha}((c_{\alpha+2}^*c_{\alpha}-c_{\alpha+2}c_{\alpha}^*)\frac{\sqrt{(\alpha+2)(\alpha+1)}}{2})-il_B^2\sum_{\alpha}((c_{\alpha+1}^*c_{\alpha}+c_{\alpha}^*c_{\alpha+1})\sqrt{\frac{\alpha+1}{2}})((c_{\alpha+1}^*c_{\alpha}-c_{\alpha}^*c_{\alpha+1})\sqrt{\frac{\alpha+1}{2}})+\frac{i}{2}l_B^2,\nn \\
\mathcal{G}_{yy}=&l_B^2\sum_{\alpha}(|c_{\alpha}|^2(\alpha+\frac{1}{2})-(c_{\alpha+2}^*c_{\alpha}+c_{\alpha}^*c_{\alpha+2})\frac{\sqrt{(\alpha+2)(\alpha+1)}}{2})-l_B^2(\sum_{\alpha}(c_{\alpha+1}^*c_{\alpha}-c_{\alpha+1}c_{\alpha}^*)\sqrt{\frac{\alpha+1}{2}})^2.
\ea

For anisotropic LL, the eigenstate $\ket{n',+}$ can be decomposed as a sum of the eigenstate of isotropic LL wave functions $\ket{n',+}=\sum_{m}(c_m\ket{m,+}+d_m\ket{m,-})$. Up to the second order, all coefficients except $c_n, d_{n-2}, d_{n+2}$ are zero and the non-zero coefficients can be written in the following way for sufficiently large LL index $n$
\ba
c_{n}=1-\frac{\Delta}{16a_1^2}, d_{n-2}=\frac{1}{4a_1}(b_2-c_3+ib_3), d_{n+2}=\frac{1}{4a_1}(b_2+c_3-ib_3).
\ea
By substituting the result, we obtain,
\ba
\mathcal{G}_{xx}=\mathcal{G}_{yy}=&
\begin{cases}
    (n+\frac{1}{2}+\frac{b_2c_3}{2a_1^2}+\mathcal{O}(n^{-1}))l_B^2, & n\ge2, \\
    (n+\frac{1}{2}+\frac{\Delta+2b_2c_3}{9a_1^2})l_B^2, & n=0, \\
    (n+\frac{1}{2}+\frac{3\Delta+6b_2c_3}{25a_1^2})l_B^2, & n=1, \\
\end{cases} \nn \\
\mathcal{G}_{xy}=&\frac{i}{2}l_B^2. \nn \\
\ea

\section{Relation between Onsager's rule and quantum geometry
\label{app:Onsager}}
In this section, we show the LL energies calculated using Onsager's rule matches \eq{eq:En2plong} up to $\mathcal{O}(n^0)$, when $\gamma=\Phi_B(\mathcal{C}_{BS})$. Let us first calculate $S_0(E_n)$. We use the same Hamiltonian we used to calculate the LL energy, which is given by \eq{eq:quad_H} with $q_1, q_2, q_3=0$. Using the energy dispersion in \eq{eq:tildeE},
\ba
S_0(E_n)=&\int_0^{2\pi}\frac{k(\theta)^2}{2}d\theta,  \nn \\
=&\int_0^{2\pi}\frac{E_n}{2a_1}(1+\frac{\Delta}{4a_1^2}+\frac{a_2}{a_1}\cos{2\theta}+\frac{a_3}{a_1}\sin{2\theta})^{-1}, \nn \\
=&\int_0^{2\pi}\frac{E_n}{2a_1}(1-\frac{\Delta}{4a_1^2}+\frac{a_2^2+a_3^2}{2a_1^2}), \nn \\
=&\frac{\pi E_n}{a_1}(1-\frac{\Delta}{4a_1^2}+\frac{a_2^2+a_3^2}{2a_1^2}),
\ea
up to $\mathcal{O}(\delta^2)$.

Next, we calculate the Berry phase of band crossing points for small $d_1, d_2$. If we expand in small $d_1, d_2$,
\ba
\Phi_{B}(\mathcal{C_{BS}})=&\int_{0}^{2\pi}s\sqrt{1-\frac{d_1^2d_2^2}{d_1^2\sin^2{\theta}+d_2^2\cos^2{\theta}}} d\theta \mod 2\pi, \nn \\
=&\int_{0}^{2\pi}s(1-\frac{d_1^2d_2^2}{2(d_1^2\sin^2{\theta}+d_2^2\cos^2{\theta})}) d\theta \mod 2\pi, \nn \\
=&-\frac{s}{2}\int_{0}^{2\pi}\frac{d_1^2d_2^2}{d_1^2\sin^2{\theta}+d_2^2\cos^2{\theta}} d\theta \mod 2\pi, \nn \\
=& -\pi s d_1d_2 \mod 2\pi.
\ea
Then if set, $\gamma=\Phi_B(\mathcal{C}_{BS})$, and plug in the results, leaving only $E_n$ at the left side, we get
\ba
E_n=&\frac{2a_1}{l_B^2}(1-\frac{\Delta}{4a_1^2}+\frac{a_2^2+a_3^2}{2a_1^2})^{-1}(n+\frac{1}{2}+\frac{sd_1d_2}{2}), \nn \\
=&\frac{2a_1}{l_B^2}(1-\frac{a_2^2+a_3^2}{2a_1^2}+\frac{\Delta}{4a_1^2})(n+\frac{1}{2}+\frac{sd_1d_2}{2}), \nn \\
=&\frac{n}{2l_B^2}(4a_1-\frac{2}{a_1}(a_2^2+a_3^2)+\frac{\Delta}{a_1}) \nn \\
&+\frac{1}{2l_B^2}(2a_1-\frac{1}{a_1}(a_2^2+a_3^2)+\frac{\Delta}{2a_1}+2sa_1d_1d_2).
\ea
This equation is equivalent to $E_{n,+}^{2band}$ in \eq{eq:En2plong} up to $\mathcal{O}(n^{0})$ since $s=1$, when $b_2, b_3, c_3>0$. Therefore, \eq{eq:En2plong} obtains higher order corrections to the Landau levels compared to the Onsager's rule, using geometric parameters.
\end{document}